\documentstyle{mn}

\input psfig
\catcode`\"=\active\let"=\"
\def\3{\ss }
\def\Stodolkiewicz{Stod\'o\l kiewicz\ }
\def\tref{$t_{{\rm rh}0}$ }

\title[A stochastic Monte Carlo approach ]
{A stochastic Monte Carlo approach to model real star
cluster evolution, II. Self-consistent models and primordial binaries}
\author[M. Giersz and R. Spurzem]
{M.~Giersz$^{1,2}$ and R.~Spurzem$^{2,3}$ \\
$^1$ Nicolaus Copernicus Astronomical Centre, Polish
  Academy of Science, ul. Bartycka 18, 00-716 Warsaw, Poland\\
$^2$ Astronomisches Rechen-Institut,
    M"onchhofstra\3e 12-14, D-69120 Heidelberg, Germany \\
$^3$ Institut f"ur Astronomie und Astrophysik, Abt. Computational Physics,
  Auf der Morgenstelle 10, 72076 T"ubingen, Germany}

\begin{document}  

\maketitle

\begin{abstract}
The new approach outlined in Paper I (Spurzem \& Giersz 1996)
to follow the individual
formation and evolution of binaries in an evolving, equal
point-mass star cluster is extended for the self-consistent
treatment of relaxation and close three- and
four-body encounters for many binaries (typically a few percent
of the initial number of stars in the cluster).
The distribution of single stars is treated
as a conducting gas sphere with a standard anisotropic gaseous model.
A Monte Carlo technique is used to model the motion of binaries, their
formation and 
subsequent hardening by close encounters, and their relaxation
(dynamical friction) with single stars and
other binaries.
The results are a further approach towards
a realistic model of globular clusters with primordial binaries
without using special hardware. We present, as
our main result, the self-consistent evolution of a cluster
consisting of 300.000 equal point-mass stars, plus 30.000 equal mass binaries
over several hundred half-mass relaxation times, well into the phase where
most of the binaries have been dissolved and evacuated from the core. In
a self-consistent model it is the first time that such a realistically
large number of binaries is evolving in a cluster with an even ten times
larger number of single stars. Due to the Monte Carlo treatment of the
binaries we can at every moment analyze their external and
internal parameters in the cluster as in an
$N$-body simulation.
\end{abstract}

\begin{keywords}
stellar dynamics -- star clusters -- numerical methods -- 
binaries, primordial
\end{keywords}

\section{Introduction}

 Dynamical modelling of globular clusters and other collisional
 stellar systems (like galactic nuclei, rich open clusters,
 rich galaxy clusters) still suffers from severe drawbacks.
 They are due partly to the poor understanding of the
 validity of assumptions
 used in statistical modelling based on the Fokker-Planck and
 other approximations on one hand, and due to statistical
 noise and the impossibility to directly model realistic particle numbers
 with the presently available hardware, on the other hand.

 Only recently a detailed comparison of the different methods
 for comparable parameter choices has been tackled (Giersz \&
 Heggie 1994a,b, henceforth GHI, GHII, Giersz \& Spurzem 1994,
 henceforth GS,
 Spurzem \& Aarseth 1996, Giersz 1996, 1998, Spurzem 1996,
 Heggie et al. 1998). They include 
 theoretical models as the
 direct numerical solution of the orbit-averaged
 Fokker-Planck equation for isotropic systems (Cohn 1980), its
 2D generalization to anisotropic models
 (Takahashi 1995, 1996, 1997, Takahashi, Lee \& Inagaki 1997) and
 rotating axisymmetric clusters (Einsel \& Spurzem 1999), 
 isotropic (Heggie 1984) and
 anisotropic gaseous models (Louis \& Spurzem 1991, Spurzem 1994)
 and direct $N$-body simulations using standard $N$-body codes
 (NBODY5: Aarseth 1985, Spurzem \& Aarseth 1996; NBODY4:
 Makino \& Aarseth 1992, Makino 1996, Aarseth \& Heggie 1998,
 NBODY6 and NBODY6++: Aarseth 1996, 1999, Spurzem 1999) or Monte Carlo
 schemes (Giersz 1996, 1998).
 All the cited work, however,
 only dealt with idealized single-mass models. There are very
 few attempts yet to extend the quantitative comparisons to
 more realistic star clusters containing different mass bins or
 even a continuous mass spectrum (Spurzem \& Takahashi 1995,
 Giersz \& Heggie 1997). 

 The results could be summarized by
 saying that in general the Fokker-Planck approximation (small
 angle two-body scattering dominates the global evolution of
 the system), the approximation of heat conduction (its energy
 transport can be treated as heat conduction in a collisional
 gas), and the statistical binary treatment (model of energy
 generation by formation and subsequent hardening of three-body binaries
 using simple semi-analytical estimates) all appear to be a
 fairly good description of what happens in $N$-body simulations.
 But there are still two basic drawbacks: (i) all
 comparisons are so far limited to rather small particle
 numbers ($N\le 64000$) as compared to real particle numbers
 of globular clusters of the order of a few $10^5$ or even
 up to $10^6$ stars. Low-$N$
 models cannot be easily extrapolated to higher $N$, since
 after core collapse a variety of different processes (close
 encounters, tidal two-body encounters, effects of the finite size
 of the stars) all vary with time scales,
 which depend on different powers of the particle number
 (see e.g. the scaling problem tackled by Aarseth \& Heggie 1998);
 (ii) during core bounce and binary driven post-collapse
 evolution an individual $N$-body simulation exhibits stochastic
 fluctuations, due to the stochastic occurrence of
 superelastic scatterings of very hard binaries with field stars
 and other binaries (three-body and four-body encounters,
 henceforth briefly ``3b'' and ``4b'' encounters).
 Although the averaged evolution of the system, understood
 either as a time average (looking for long post-collapse times)
 or as an ensemble average (averaging statistically independent
 single $N$-body models), is reproduced well by the
 theoretical models based on the above assumptions, the
 {\sl individual} evolution of a stellar system, even with
 a relatively large particle number, might not be exactly matched
 at any instant. The most recent collaborative experiment in
 this area (Heggie et al. 1998) gives a good overview:
 all methods do agree fairly well, but variations of quantitative
 results of some 10 or 20 \% and some scaling problems, which
 are not exhaustively examined, have to be tolerated.

 Such considerations led to the construction of special-purpose
 computers for direct $N$-body simulations (Sugimoto et al. 1990,
 Makino et al. 1997, Makino \& Taiji 1998) and considerable efforts to improve
 the highly accurate $N$-body simulation software used (see
 e.g. Aarseth 1996, 1999, Spurzem 1999, but also the alternative approach
 by KIRA mentioned e.g. in Makino \& Taiji 1998,
 McMillan \& Hut 1996, Portegies Zwart et al. 1998).
 But there is an elegant alternative way to generate models of
 star clusters, which correctly reproduce the stochastic features
 of real star clusters, but without really integrating
 all orbits directly as in an $N$-body simulation. These so-called Monte Carlo
 models were first presented by H\'enon (1971, 1975, Spitzer 1975) and later
 improved by \Stodolkiewicz (1982, 1985, 1986) and in further
 work by Giersz (1996, 1998). The basic idea is
 to have pseudo-particles, whose
 orbital parameters are given in a smooth, self-consistent potential.
 However, their orbital motion is not explicitly followed; to
 model interactions with other particles like two-body relaxation
 by distant encounters or strong interactions between binaries and
 field stars, a position of the particle in its orbit and further
 free parameters of the individual encounter are picked from
 an appropriate distribution by using
 random numbers. For a more detailed description see
 the cited papers and Sect. 2 below.

 In the past it proved to be very delicate to properly tailor
 flexible and versatile pure Monte Carlo models.
 Stochastic fluctuations of the parameters 
 of $N$-body systems with large particle
 numbers (say $N\ge 10000$) are mainly due to binary effects
 (formation by 3b encounters, superelastic scatterings
 with field stars, or, in the case of the presence of many
 primordial binaries, binary-binary encounters, see e.g.
 Heggie \& Aarseth 1992, henceforth HA92), 
 not so much due to the inherent fluctuations
 resulting from the discrete nature of the particle system
 (which should decay approximately with $\sqrt{N}$).
 Therefore the idea appeared only to model the binary population
 by a Monte Carlo technique, above a background of single stars,
 which are treated by a standard theoretical model. Takahashi
 \& Inagaki (1991) published a similar approach for the case
 of an isotropic Fokker-Planck model, but without following
 individual binary's orbits and their relaxation interaction
 with themselves and other single stars. In an earlier
 approach Inagaki \& Hut (1988) simulated the stochastic evolution
 of a binary population including a distribution of
 binary orbits. However, their
 background single star cluster was only approximately modelled,
 and their binaries were assumed to have purely radial orbits,
 and dynamical friction was treated approximately (as opposed to
 a full Monte Carlo model, which models explicitly the cumulative effect
 of many small angle encounters 
 in order to generate proper relaxation and dynamical friction effects).

 Here we want to extend the stochastic treatment of
 binaries in an anisotropic gaseous model for the single
 stars, as introduced in Spurzem \& Giersz (1996, henceforth Paper I),
 for the self-consistent inclusion of
 many primordial binaries in a globular cluster model (which is
 yet a simplified model, with regard to its neglect of any 
 finite size effect of the stars and its assumption that all
 stars have equal masses). It is very clear now from the
 observational (Hut et al. 1992) as well as the theoretical viewpoint of
 star formation that globular clusters do contain of
 the order of 5-15 \% of binaries, which means that at the
 time of their formation the binary fraction had to be even larger,
 because many soft binaries are subject to disruption by close
 3b and 4b encounters. Nevertheless theoretical modelling
 of the dynamical evolution of even idealized large star clusters containing
 many binaries has not advanced very much in the last years. 

 Pioneering studies by McMillan, Hut \& Makino (1990, 1991),
 McMillan \& Hut (1994) using a direct N-body integrator with regularization
 techniques for close binaries (NBODY5: Aarseth 1985) used some
 1000 particles of equal mass only (binary membership of the order of
 10-20 \%). A similar work by HA92 provided
 in some more detail went up to $N=2000$, with less than 10 \% of binaries.
 The results of such models was that the details of the initial
 distribution of primordial binaries, at least for such small
 total particle numbers, strongly influences the global dynamics
 of the star cluster, which is quite understandable, since usually
 the binaries' total binding energy is much larger than the binding
 energy of the entire star cluster.

 On the high-$N$ branch Fokker-Planck modellers tried to incorporate
 appropriate average cross-section for 3b and 4b
 encounters in their simulations, as was done so successfully
 for the 3b case (Lee et al. 1991, Giersz \& Spurzem 1994). 
 In a pioneering study Gao et al. (1991, henceforth GGCM91) published the first
 self-consistent model of an N=330.000 star cluster with as
 many as 30.000 binaries and showed how gravothermal collapse
 (and subsequent oscillations) were delayed by a large factor in
 time due to previous ``burning'' of primordial binaries. 
 In their model the close binary encounters, however, had to
 be treated in a very approximate way, as we will discuss later.

 Unfortunately the direct $N$-body modelling of a case like the one in the
 GGCM91 paper is hardly possible today even with the fastest
 special purpose computers, because both the
 scaling of the computational time for the general $N$-body
 problem is prohibitive and as well as the regularization of the close
 binaries downgrades the performance very much. In this paper
 we use the hybrid Monte Carlo code presented
 in Paper I, which combines an anisotropic gaseous model
 for the single star component with a Monte Carlo stochastic treatment of
 many individual binaries. In our largest models we are able
 to follow 300.000 stars and 30.000 binaries as in the paper
 by GGCM91.
 Our model is entirely self-consistent in the way that
 both the binaries and single stars are subject to their own and
 the other component's gravitational field. The binaries are treated
 individually with their orbital and internal parameters (eccentricity, 
 binding energy, semi-major axis). Relaxation of the binaries with 
 themselves and the single stars as well as close 3b and 4b
 encounters are treated in a direct, but stochastic way using approximate
 cross sections for the latter and the standard Monte Carlo procedure
 (H\'enon 1971, \Stodolkiewicz 1982, 1986, Giersz 1998) for the relaxation.
 We use the same probability distribution for the hardening of the
 harder binary by close 4b encounters as GGCM91 (originating from
 Mikkola 1983a, b, 1984a, b); different to them, and more
 realistically, we compute
 the probability for a close 4b encounter to occur in a self-consistent
 way from the local density distribution of our binaries.
 We can also in detail follow the
 evolution and compile balances of all the reaction products
 in any close encounter (singles and binaries).

 At this place some more sophisticated models of the evolution of
 a large binary population should not go unmentioned, i.e. the
 Hut et al. (1992) model with a very detailed study of finite-size
 effects and using unequal masses, similarly Sigurdsson \& Phinney (1995),
 but both using a static background of single stars. Another interesting
 subject related to this study is the question of the evolution of
 binary parameters in young star forming clusters in the galaxy,
 which was studied by Kroupa (1995) by using NBODY5 for direct $N$-body
 simulations. Our model will be soon further developed to incorporate
 multi-mass systems, which will render it applicable
 for all such purposes.

 In the next section we will describe only those features of the model
 which have been added as compared to Paper I. For the
 treatment of the ordinary relaxation process and the
 close 3b encounters we refer to Sect. 2 of Paper I.

\section{Self-Consistent Monte Carlo treatment of binaries}

\subsection{General remarks}

In Paper I stochastic binaries were followed in their motion
and interactions
(two-body relaxation or dynamical friction, and close 3b
encounters) with the single stars. The only feedback to the
single stars, which were treated by the gaseous model, was the
kick heating due to superelastic 3b scatterings. For
Fokker-Planck models Takahashi
\& Inagaki (1991) published a similar approach
but without
any motion of the binaries in the system. In an earlier conference
proceedings
Inagaki \& Hut (1988) reported a stochastic model
of a binary population including binary orbits. However, their
background model cluster was only approximately modelled,
the binaries were assumed to have purely radial orbits,
and dynamical friction was treated approximately (as opposed to
a full Monte Carlo model, which models explicitly the cumulative effect
of many small angle encounters
in order to generate proper relaxation and dynamical friction effects).
To our knowledge their approach was never continued nor published elsewhere.

In order to improve our models to make them applicable
for systems with many (primordial) binaries one has to cope with
a number of complications. First, the energy balance is changed
due to 4b (binary-binary) close encounters, and 
binary-binary small angle relaxation has to be taken into account
according to H\'enon's Monte Carlo method (H\'enon 1971). Second, the
energetic feedback effect of binary-single star relaxation cannot
be neglected any more for the energy balance of the single stars. Third,
binaries move in the system due to relaxation and close encounters, and so
they change the total mass and potential of the entire system. Another
complication occurs due to the losses and gains of single stars by 
binary formation and disruption of
binaries in close 4b encounters.
All such adjustments
generate a feedback via the potential gradient in Euler's equation
to the single stars. The first points were relatively easy to implement:
we adopted the approximate 4b cross sections given by GGCM91
(1991) as a working model, to be tested and improved later; the 
treatment of binary-binary relaxation is similar to
Giersz (1998), except that the local binary density
is estimated here from the smoothed out binary
mass (see below) via Poisson's equation. 
The second and third point, however,
appeared to be extremely difficult to cope with. The problems were
due to the very different nature of the gaseous and Monte Carlo
treatment for the two components. The gaseous model approximates
density and potential by smooth functions using Newton's method
to find the iterative solution at the next time step. In contrast
to this the stochastic binaries may experience rather quick and
unsteady changes of their positions, velocities, and binding energies. 
Incorporating
such changes in a naive way into the gaseous model (just adding up
energy and mass balances per time step and applying them as a 
local source or sink term in the continuity and energy equations) yielded
catastrophic failure. Thus we had to invent a number of measures
to soften the transitions and links between the two systems
without spoiling the physically correct treatment of the
combined system and still preserve
the advantages of the hybrid method. The physical justification
is that the gaseous model is to be interpreted as an ensemble
average over the evolution of many independent individual
star clusters. So, when applying mass or energy changes we have
to properly average out stochastic variations of the binaries.
Our resulting code is technically
fine-tuned and, frankly speaking, rather fragile according to our experience.
However, with this paper we want to publish results obtained with
a final code version, which we believe to produce physically correct
results, in an efficient way, and without any unexplained or unphysical
procedures. Due to difficulties described above we think it is justified to
give in the following subsections
some rather technical information how
the implementation is done.

\subsection{Binary mass distribution, relaxation, and heating processes}

Our main idea to cope with the mentioned problems is to 
include the mass of each binary in the gaseous model as a smooth
orbital mass distribution, computed by using the orbit parameters
of the binary, which are known from the Monte Carlo model. 
For the stochastic treatment, the position of a binary is still
determined by the appropriate picking procedure (Giersz 1998), and
if we plot our results we give such positions of binaries. However,
in Euler's equation in the gaseous model a smoothed out mass is
used. Thus local potential fluctuations are reduced, since they
now only reflect the change of the orbital parameters of the binary,
not the stochastically picked individual binary position. 
So from the viewpoint of the gaseous model a binary appears as a
smooth mass distribution, extending from its minimum
to maximum orbital radius in the cluster. The mass distribution in the orbit
is weighted according to $dr/v_r$.
Thus the total binary
mass $M_b(r)$ contained in a sphere of radius $r$, which enters into the
force equation of the gaseous model is obtained from

\begin{equation} 
M_b(r) = \sum_j M_{bj}(r) 
\label{e1}
\end{equation} 
where $j$ is an arbitrary index counting the binaries and $M_{bj}(r)$
is computed by integrating the following equations:

\begin{eqnarray}
M_{bj}(r) && = 0 \ \ {\rm for} \ \ r<r_{\rm min}(E_j,J_j) \nonumber \\
M_{bj}(r) && = m_{bj} \ \ {\rm for} \ \ r>r_{\rm max}(E_j,J_j)
\label{e2}
\end{eqnarray}
with the total mass of the $j$-th binary $m_{bj}$; for 
$r_{\rm min}<r<r_{\rm max}$ we use

\begin{equation}
M_{bj}(r) = {m_{bj}\over P_j} \int\limits_{r_{\rm min}}^r 
 {dr \over v_r}
\label{e3}
\end{equation} 
with the orbital period

\begin{equation}
P_j = \int\limits_{r_{\rm min}}^{r_{\rm max}} {dr\over v_r} \ . 
\label{e4}
\end{equation} 
Note that 

\begin{equation} 
v_r^2 = {2\over m_{bj}}\bigl(E_j - \Phi(r)\bigr) - 
  {J_j^2 \over m_{bj}r^2}
\label{e5}
\end{equation} 
depends itself on $r$. Since a standard 
numerical integration of the integrals in
Eqs. \ref{e3} and \ref{e4} 
is very difficult near the turning points of the orbit 
($v_r\rightarrow 0$), we use H\'enon's Monte Carlo procedure  (H\'enon 1971) to
pick between 50 to 500 radial positions in the orbit (depending on how
eccentric it is), compute $v_r$ for each radius,
and approximate the integrals in Eqs. \ref{e3} and \ref{e4} by
a summation over these stochastically picked radii.
The gravitational potential of the binary component is then
computed in a standard way from the mass distribution
using the approximation of spherical symmetry
($\Phi(r)\rightarrow 0$ for $r\rightarrow\infty$).
A slight inconsistency enters here, because the new potential is not
yet known in Eq. \ref{e5} for the computation of the new 
binary mass distribution (see Giersz 1998).

One more step is necessary to make the gaseous model compatible with
many stochastic binaries. If its time step
in high density phases becomes as small as some fraction of the central
relaxation time, close encounters and relaxation encounters
with sufficiently big deflection angles cause substantial
heating and cooling in the stellar core. Note that the
Monte Carlo model simulates the cumulative effect of small
angle gravitative encounters by representative encounters, whose
deflection angle does not need to be very small.
Applying the energetic effect of such encounters immediately in
the gaseous model component for the single stars would disrupt
the stability of the system,
because the time scale of the changes becomes much shorter than the
local relaxation time. Therefore we define a reservoir of energy,
which is applied to the single stars with a maximum rate of
$E_c/t_{\rm rx,c}$, where $E_c$, $t_{\rm rx,c}$ are the
core energy and the central relaxation time,
which is smoothly distributed in the inner 20\% of
the total mass of the cluster, with a power-law cutoff.
We interpret this as a simplified model of how
the sub- or suprathermal reaction products
themselves relax with the other cluster members.
For the sink and source terms of the mass we use a similar
procedure.
With these measures the gaseous model code
is able to cope with the induced mass and energy variations of
an arbitrary number of binaries. 

Heating terms contain contributions
from the following processes:

\begin{itemize}
\item
 energy feedback of relaxation encounters
 between singles and binaries to the single stars. It is easily
 accomplished, since for each relaxation event in the standard 
 Monte Carlo procedure we know the deflection angle and the
 energies of the reaction products after the encounter. It is
 checked whether the single star can escape as a result of
 its relaxation encounter with the binary. If this is not the
 case the energy it gained is added as a heating (or cooling) source
 to the gaseous model equations. 
 The relaxation procedure itself is the same as
 described in Paper I, but now the energy feedback to single stars
 is not considered negligible.

\item
 heating due to close 3b encounters. This is the classical
 heating by three-body binaries, which was first implemented in
 a steady approximation by
 Heggie (1984) and Bettwieser \& Sugimoto (1984). Its Monte Carlo realization
 depends on an assumed probability for an encounter to occur and
 a probability distribution of $\Delta E_b$, the change of binary binding
 energy (see Spitzer 1987, and detailed description in Paper I). 
 Again, the heating is only applied if the single star reaction
 product of a close encounter does not escape. The treatment of this
 process is equal to Paper I.
 
\item
 heating due to close 4b encounters. With many binaries in
 the system, the
 probability of close binary-binary encounters cannot be neglected any more.
 The problem is very difficult to solve. However, as in the
 case of 3b encounters, for hard binaries a typical event
 may be defined (at least in the case where all single stars have
 equal mass). The typical close encounter between two
 hard binaries generates a ``hardening'' (gain in
 binding energy) of the harder binary of the two and a disruption of
 the other one into single stars. Depending on the initial binding energies
 the single stars created may escape. If not they
 contribute to the heating term described above. 
 To determine the probability for a close 4b encounter we
 sort all binaries radially according to their most recently
 picked position in their orbits and group them into pairs. For each pair
 the probability is computed and compared to a random number,
 insofar this is again the standard Monte Carlo treatment (Giersz 1998).
 The probability $P_{bb}$ for a close encounter to occur is determined,
 however, here by using an $n\sigma v$-ansatz:

\begin{equation}
P_{bb} = n_b \pi R_b^2 v_{\rm rel} 
 \cdot \Bigl(1 + {2 G m_b \over R_b v_{\rm rel}^2}\Bigr) 
\label{e6}
\end{equation} 
 where $n_b = \rho_b/m_b$ is the local binary number density (determined
 from the smoothed out binary mass via a spherically symmetric Poisson's
 equation), $m_b$ the binary mass, $v_{\rm vel}$ the relative velocity
 of the two selected binaries, and $R_b$ is 2.5 times the semi-major
 axis of the harder binary (\Stodolkiewicz 1986). In his paper a different
 ansatz was given for $P_{bb}$, but we checked that by order of magnitude
 it provides the same results. By comparing $P_{bb}$ with a random
 number it is decided whether an interaction takes place. If so,
 we use the probability distribution given by GGCM91
 (taken from earlier work by Mikkola 1983 a, b, 1984 a, b)

\begin{equation}
G(y) = {49\over 4} y \Bigl(1 + {7\over 2} y^2 \Bigr)^{-11/4} 
\label{e7}
\end{equation} 
 which gives the fractional translational or recoil energy $E_r$
 released in the encounter, defined by

\begin{equation}
y = {E_r \over E_b + E_b^\prime} 
\label{e8}
\end{equation} 
 ($E_b$, $E_b^\prime$ being the two binding energies). The increase in
 binding energy of the harder binary is thus

\begin{equation}
\Delta E = y (E_b + E_b^\prime) + E_b^\prime
\label{e9}
\end{equation} 
 since the softer binary has to be disrupted. One quarter of
 the recoil energy $E_r$ is assumed to be carried off by the remaining
 binary, three quarter are distributed randomly between the two stars. 
 Such a procedure is very approximate, and it neglects certain other
 channels of binary-binary reactions, such as the formation of stable
 hierarchies with three or more particles (compare Aarseth 1999).
 While our Monte Carlo model in future work will be capable to 
 include more details of 4b encounters, for the model presented here
 we decide to use the same reaction processes as GGCM91, for a 
 comparison of our results with theirs. 

 We use the Monte Carlo rejection technique
 to pick random values from a distribution $f(z) = z^{3/4}$; in
 case a value is accepted we determine

\begin{equation}
y = \sqrt{{2\over 7} {1-z\over z}}
\label{e10}
\end{equation} 
 where $z$ is the original random number. This provides the above
 probability distribution $G(y)$. By checking the translational
 energies of the reaction products (one binary and two singles) we
 determine whether they escape or not. If the single stars do not
 escape, their translational energy is added to the heating source for
 the single star component, and their mass is added to the single
 stars as well. Again, as described above for the energy, the
 mass is taken away from or added to the single star component smoothly
 in a region defined by the 20\% Lagragian radius with a power-law
 cutoff.

 \end{itemize}

\subsection{Adjusting the masses and energies}

 The following processes change the masses contained in the single
 or binary components and are properly accumulated and taken from
 or given to the single stars in form of a source or sink term in
 the continuity equation:

\begin{itemize}
\item formation of three-body binaries; this process is treated
 as in Paper I, but now the mass loss for the single stars is
 properly accounted for as a sink term in the gaseous model equations;

\item generation of two single stars as a reaction product from
 a 4b encounter, which do not escape; this is accounted for
 in a mass source term in the continuity equation;

\item mass loss of singles and binaries by relaxation, close
 3b and 4b encounters. This is mass loss to the entire
 system which causes its total binding energy to decrease.

\item finally, there is an ongoing motion of binaries in the
 system, due to relaxation with each other and with single stars
 (dynamical friction) and due to 3b and 4b encounter kicks. 

\end{itemize}

 To check total energy conservation we use a balance equation with 
 respect to the total energy $E_{\rm tot}$ of the single stars only:

\begin{equation}
2E_{\rm tot} = \mu v^2 + \mu  (\Phi_s + 2\Phi_b)
\label{e11}
\end{equation} 
 where $\Phi_s$, $\Phi_b$ are the local potentials from singles and binaries,
 respectively, and $v$ is the 3D velocity dispersion of the single stars.
 Note that the factor of 2 in front of the binary potential here occurs,
 because the binaries provide an external potential and do not belong
 to the self-gravitating single star system.
 To simplify the following discussion we consider the above 
 as an energy balance
 for one discrete radial shell in our model. To get the total energy
 this has to be integrated (or in the discrete numerical model summed)
 over the volume. In this sense  $\mu  = \rho\Delta V $ is the mass of
 single stars contained in one radial shell. Any changes in our model
 cause a change 

\begin{equation}
2\delta E_{\rm tot} = v^2 \delta \mu  + \delta \mu  (\Phi_s + 2\Phi_b) + 
     \mu \delta v^2 + \mu  (\delta \Phi_s + 2 \delta  \Phi_b). 
\label{e12}
\end{equation} 
 Checking the total energy of the single stars in our model means
 book-keeping of all processes which create changes in one of the
 quantities of the above equation.
 
\begin{itemize}
\item
 first, there is a change of $\Phi_b$ due to relaxation, 
 close 3b and 4b kicks of the binaries,
 binary escape, and escape of single stars which is
 caused by a binary-binary encounter. All such processes
 lead to changes in the orbital mass distribution of the binaries,
 but not to direct changes of the state of the single stars;
 thus we have

\begin{equation}
2\delta E_{\rm tot} = 2 \mu  \delta \Phi_b.
\label{e13}
\end{equation} 
 If however, single stars escape from the system which belonged
 as single stars to the cluster before (this is only possible
 by close 3b encounters between a binary and a single
 star) we have an energy change of

\begin{equation}
2\delta E_{\rm tot} = 2 \mu  \delta \Phi_s.
\label{e14}
\end{equation} 
 because the adjustment of the self-gravitating system proceeds
 such that $\Phi_s\delta\mu = \mu\delta\Phi_s$.

\item
 second, there is mass loss of single stars due to recoil ejections
 in close 3b encounters. From our numerical experiments
 we find that in the case of such an adjustment the system reacts
 with $\delta v^2 \ll 1$ and $\Phi_s\delta\mu = -\mu\delta\Phi_s$,
 i.e. the thermal structure is not changed, and the change
 in the mass distribution is fully represented by the change
 of potential energy. Neglecting the first term and using the
 second identity in Eq. \ref{e12} yields:

\begin{equation}
2\delta E_{\rm tot} = v^2 \delta \mu + \delta \mu  
       (\Phi_s + 2\Phi_b) + \mu \delta \Phi_b.
\label{e15}
\end{equation} 
  
\item finally, there are two processes, which shift mass between
 singles and binaries, but do not change the total mass distribution,
 i.e. $\Phi_s + 2\Phi_b$ remains constant. They are formation of binaries
 by 3b encounters and disruption of binaries by 4b encounters
 leading to two non-escaping single
 stars. Here again the above equation seems
 to be the proper ansatz. However, we observe that in such a case the
 adjustment of the gaseous model proceeds in a way that

\begin{equation}
2\delta E_{\rm tot} = \delta \mu v^2 + \delta \mu (\Phi_s +
  2\Phi_b) + {2\over 3} \mu \delta \Phi_b.
\label{e16}
\end{equation} 
 This follows from the fact that the mass given to or taken from the single stars
 is not localized but distributed smoothly inside the central 20\% of total
 mass, which introduces an additional shift in both the binaries' and singles'
 potential. Again from numerical experiments we find a good approximation of
 the identity $\delta\Phi_s/2 = - \delta\Phi_b/3$. 

 \end{itemize}

 The reader should keep in mind, however, that
 all the described rather complicated details about the energy balance
 for the different processes do only matter for the purpose of
 checking the conservation of energy in the entire calculation.
 They have no influence on the physical events and progress of the
 model itself. But they were essential for us in the process of
 debugging the method and finding proper descriptions for all
 processes. In one of our models (Gao's run) the total binding
 energy of the binaries exceeds the binding energy of the
 system by several orders of magnitude, and this again is some
 orders of magnitude larger than the total energy error we find after
 a very long integration time. This is only possible if our procedure
 is correct. 

 \subsection{Final remarks}

 Having described the zoo of different processes and balances in the
 previous subsections should illustrate to the reader, that in a 
 stochastic binary model like ours we have a detailed and individual
 book-keeping of all these processes. It ensures that we have practically the
 same amount of information as in a large $N$-body simulation with
 primordial binaries. The underlying additional approximations 
 here are just spherical
 symmetry, no relaxation processes other than standard two-body, and
 the cross sections for close 3b and 4b encounters. The latter
 will be improved to a detailed numerical three- and four-body integration
 in the future. $N$-body simulations with primordial binaries have
 been published by McMillan, Hut \& Makino (1990) and HA92.
 In no case was the binary number larger than 150, and the total
 star number did not exceed 2500. We use the models of HA92
 as a template for our models, to check whether there are differences
 and how significant they are. They (around 1992) needed
 2000 h CPU time on a 10 Mflop workstation, hence
 the problem requires 72 Tflop. So it is a major computational job still
 today, though not a question of months. But our real interest stems
 from globular clusters, which may start with initial data like
 the ones used in the Fokker-Planck models of GGCM91,
 using 300.000 single stars and 30.000 primordial binaries. 

 It is
 questionable, whether this job can be done in a direct simulation
 even by using a Petaflops computer. We see the justification in our
 efforts to create stochastic and Monte Carlo models of star clusters
 with many binaries in the fact, that our model is with a computational
 effort of two weeks on a Pentium II PC, able to provide a full  
 model for the GGCM91 case. In excess
 it yields a wealth of more details of all binary related processes
 and the full stochasticity of the binaries {\em as it could
 only be deduced otherwise from an enormous $N$-body simulation}.   
 Note that some of the approximation used in our model will be
 released in the near future without fundamental difficulties, 
 such as the cross sections for 3b and 4b interactions.
 On the contrary there are assumptions, which are rather fundamental 
 for our model, and not straightforward to remove. They are spherical
 symmetry and the assumption that only two-body relaxation by
 small angle encounters is important. Such a treatment
 of relaxation may be justified, because 
 collective processes are less important in large $N$-body systems. 
 Regarding the assumption of spherical symmetry,
 all Monte Carlo models so far employed it, because
 it makes the potential computations much quicker and it always leads to
 purely deterministic orbits. This assumption is a much more severe
 restriction in the moment, because a treatment of axisymmetric
 systems would require a
 not straightforward reformulation of all Monte Carlo methods 
 (similar as in the
 case of Fokker-Planck models, Einsel \& Spurzem 1999).

 One final note: the reader making comparisons between the results presented
 in the following sections as well as results by HA92,
 Hut, McMillan \& Romani (1992), and GGCM91 should keep in
 mind that all results (except for the Fokker-Planck models of GGCM91) are
 individual representations of a statistical ensemble of solutions;
 all energy generation events for example occur randomly and the
 outcome of such close 3b and 4b encounters is treated here purely
 stochastic. So one should not expect a perfect match between all
 models, but an agreement of global and averaged parameters of the
 simulated systems. By using different random number sequences for
 our models, however, one can generate a statistical spread of the results,
 which is comparable to the differences between Monte Carlo
 models here and direct $N$-body results in HA92.

 In the following we present different set's of test models by applying our new
 method, which we call Monte Carlo runs and Heggie's runs, due to
 the aim to compare with corresponding other works.
 Finally, we model a star cluster with
 300.000 single stars and 30.000 binaries, called Gao's runs,
 which has no direct
 counterpart in the direct simulations, so we have to rely on the
 capabilities of our code and the comparison with the more approximate
 GGCM91 models.

\section{Results and Models}

\subsection{Method}

 The dynamical equations of the gaseous model
 were discretized on a logarithmically
 equidistant mesh of 200 points ranging over eight orders of
 magnitude and solved by an implicit Henyey-Newton-Raphson scheme
 (see GS). More details on the Monte Carlo description of our
 binaries can be found in Paper I, the previous section and
 the cited literature. All results presented in the paper
 are (if not explicitly stated otherwise) given in standard
 units, i.e. $G=M=1$ and $E=-1/4$ (Heggie \& Mathieu 1986), where
 $M$ and $E$ denote the initial total mass and translational
 energy of all members of the
 cluster (including single stars and binaries). The total
 energy used for this normalization does not include the internal
 (binding) energy of the binaries.

 \subsection{Monte Carlo Runs}

 \begin{figure}
 \psfig{figure=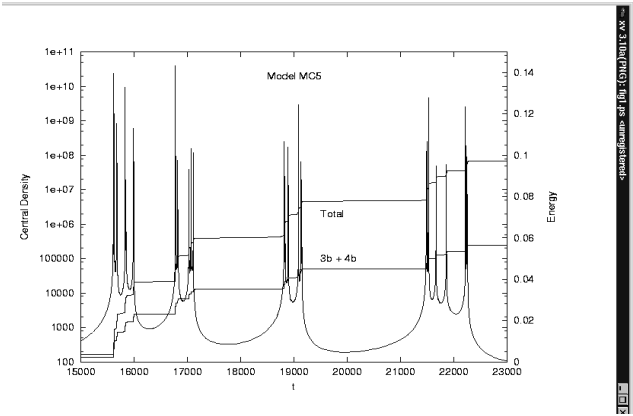,height=5.5cm,width=8.5cm,angle=-90}
 \caption{Model MC5, central density, total heating
 of the system, and heating due to 3b and 4b close encounters only,
 as a function of time.}
 \label{f19}
 \end{figure}

 \begin{figure}
 \psfig{figure=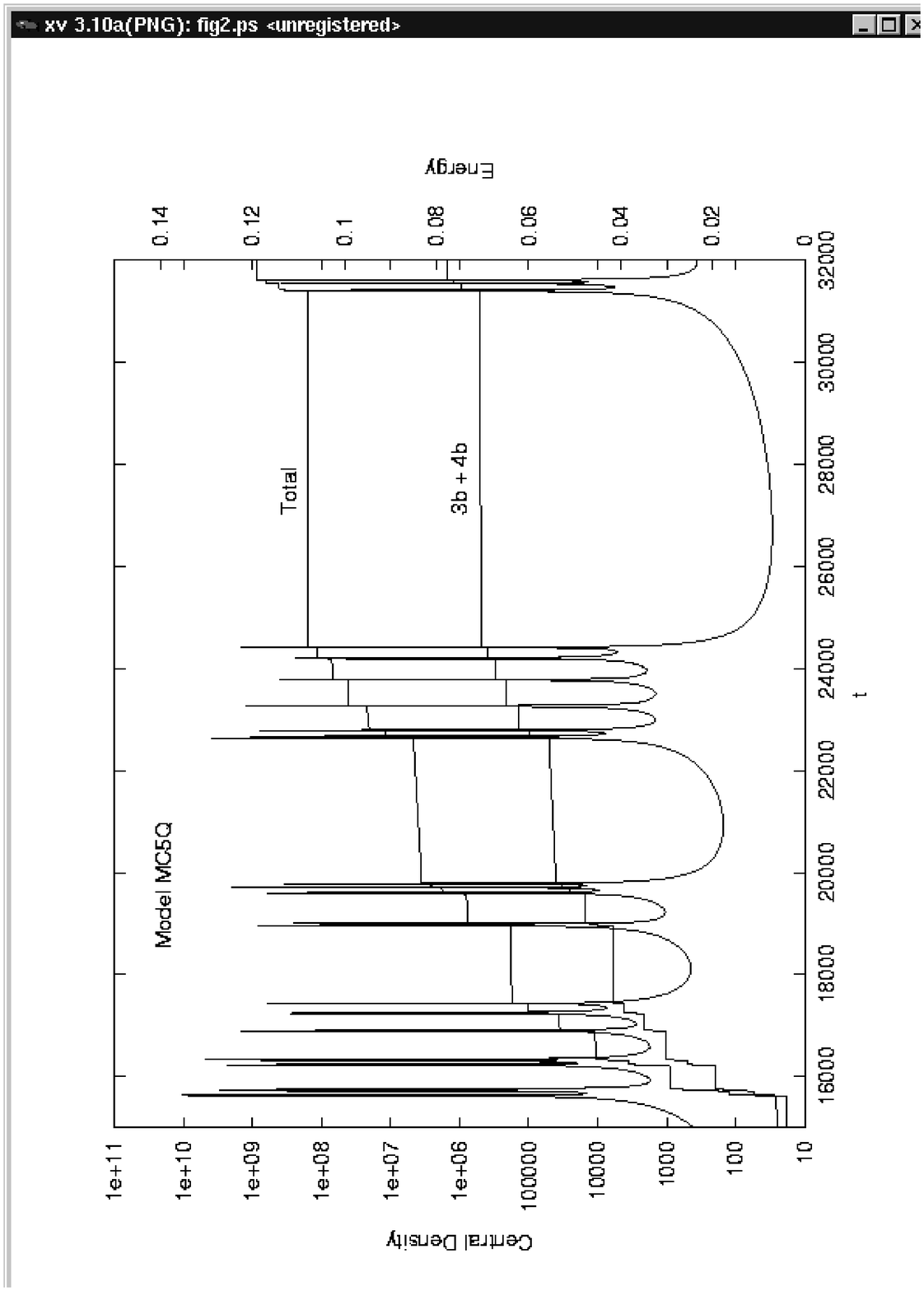,height=5.5cm,width=8.5cm,angle=-90}
 \caption{Model MC5Q, as Fig. \ref{f19}}
 \label{f20}
 \end{figure}

 \begin{figure}
 \psfig{figure=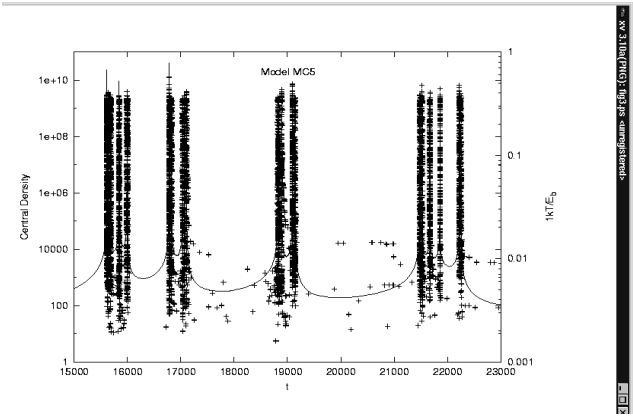,height=5.5cm,width=8.5cm,angle=-90}
 \caption{Model MC5, central density plus crosses
 for each close 3b and 4b encounter as a function of time. The abszissa
 of the crosses is chosen according to $1kT/E_b$,
 in order to illustrate the correspondence between events and fluctuations
 in the density curve.}
 \label{f21}
 \end{figure}

 \begin{figure}
 \psfig{figure=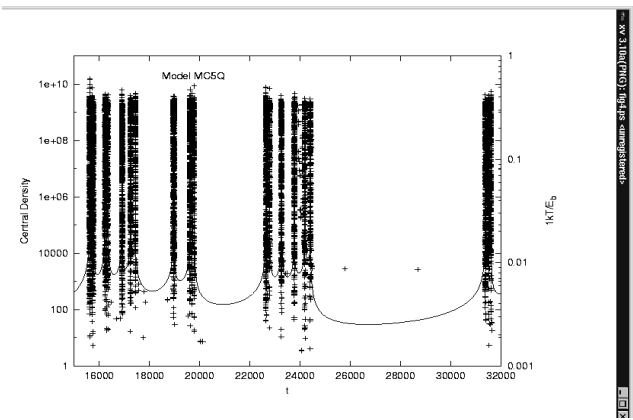,height=5.5cm,width=8.5cm,angle=-90}
 \caption{Model MC5Q, as Fig. \ref{f21}}
 \label{f22}
 \end{figure}

 To check the reliability of the stochastic Monte Carlo model we studied
 models of a system consisting of $10^5$ single stars without
 any primordial binaries. Binaries in these runs are
 created only due to the dynamical 3b interactions between single
 stars. In one run, labelled MC5, binaries only harden due to interactions
 with single field stars, but in another run,
 labelled MC5Q, they also interact between
 themselves. Both runs are compared to full Monte Carlo models 
 discussed by Giersz (1998).

 The evolution of the central density together
 with the different sources of heating of
 the gaseous system are presented in Figs. \ref{f19} and \ref{f20} for models
 MC5 and  MC5Q, respectively. The curve labelled ``3b + 4b'' shows the heat
 transferred to  the single stars (gaseous component) due to binary
 hardening (connected with  their interactions with field single
 stars and other binaries). The curve labelled ``Total'' shows the total
 heating of the gaseous system including ``3b + 4b'' contributions (see
 above) plus an additional heating or cooling
 due to the motion of binaries in the
 system. Note that binaries from the point of view of the gaseous model form
 an external system. Their motion caused by small
 angle two-body interactions with single stars (relaxation process) induces
 changes of the total potential felt by the single stars.
 The total heating is always larger than that connected only with 
 the strong binary
 interactions (``3b + 4b''), 
 because the binaries preferentially lose their transational
 energy in the relaxation process and sink to the centre of the system.
 Therefore the relaxation interaction with the binaries is always, in
 an averaged sense, a
 heating (the ``Total'' curve is always above the pure
 ``3b + 4b'' curve). Note that individual relaxation encounters
 between binaries and single stars may lead to cooling also.
 This is in agreement with the expectation from the physical
 picture of mass segregation. 

 Large scale quasi-periodic
 oscillations are the most
 pronounced features of these two figures. They span several orders of magnitude
 in central density and have strongly varying periods.
 A characteristic
 feature of these oscillations is the lack of binary energy generation
 during the long phases of maximum expansion. It is widely accepted that this
 together with a temperature
 inversion (during these phases) are the most important signatures of
 gravothermal oscillations (Bettwieser \& Sugimoto 1984, McMillan \& Engle
 1996). 

\begin{figure*}
\vskip 13truecm
\includegraphics{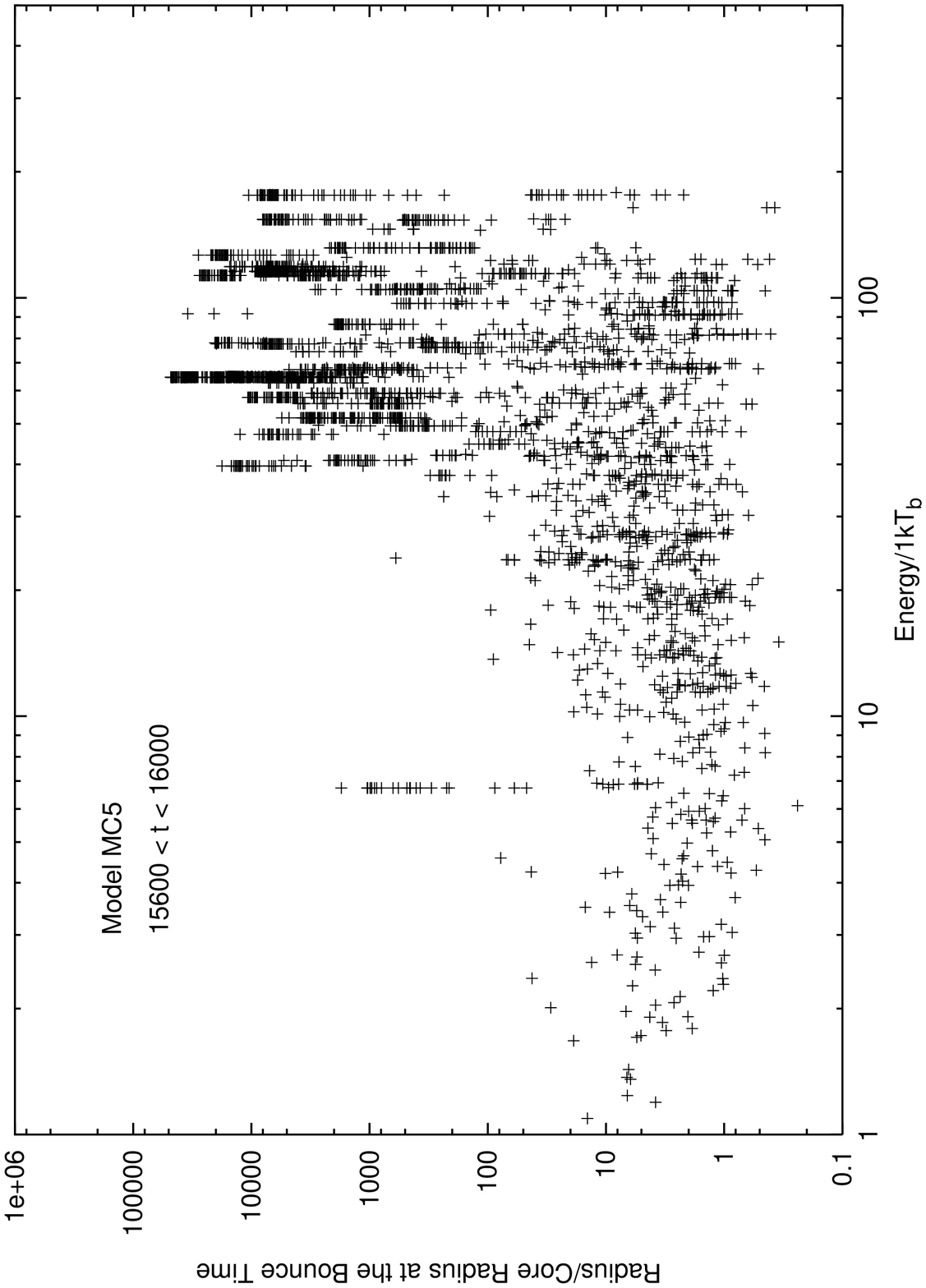}
\includegraphics{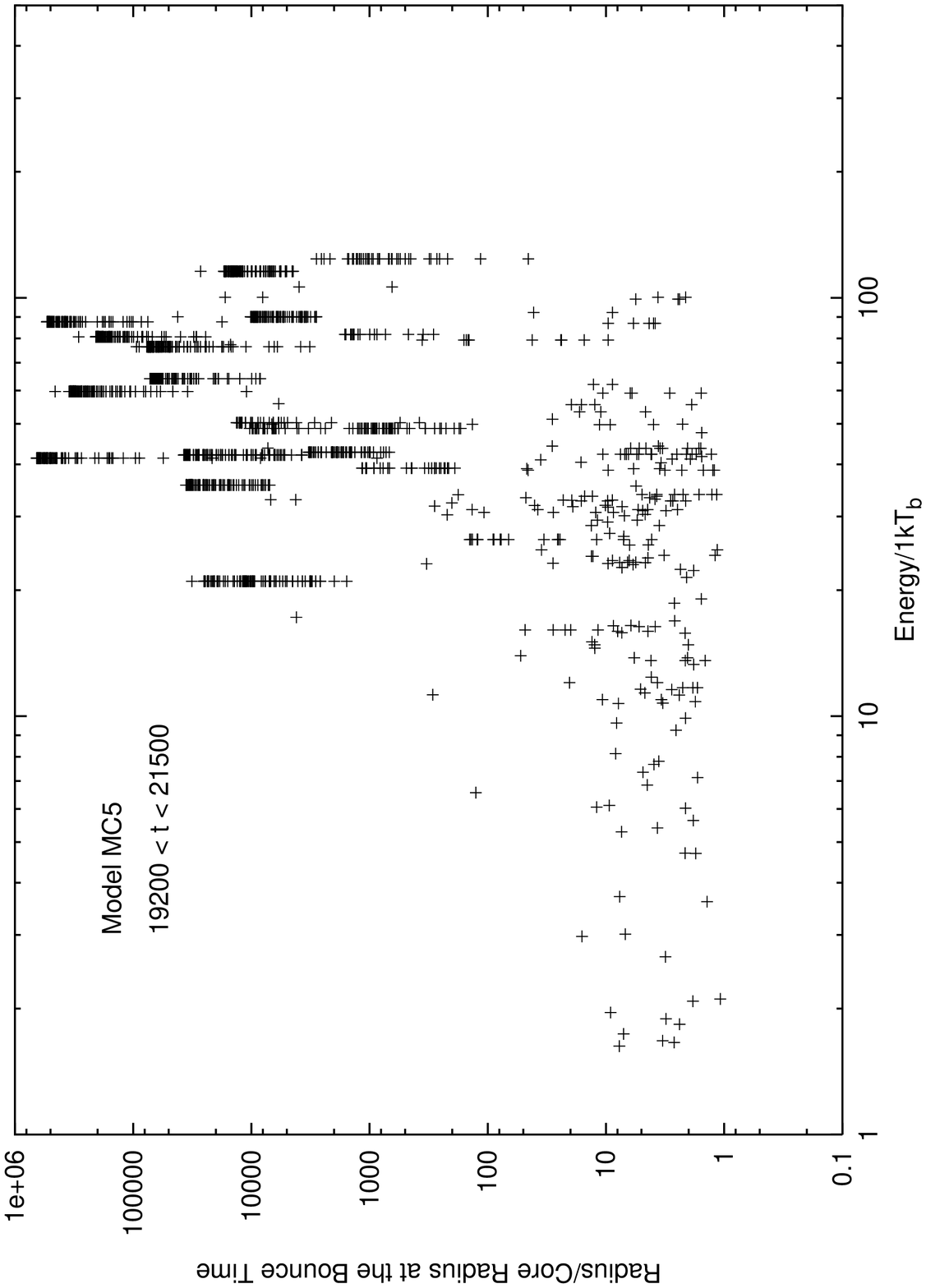}
\includegraphics{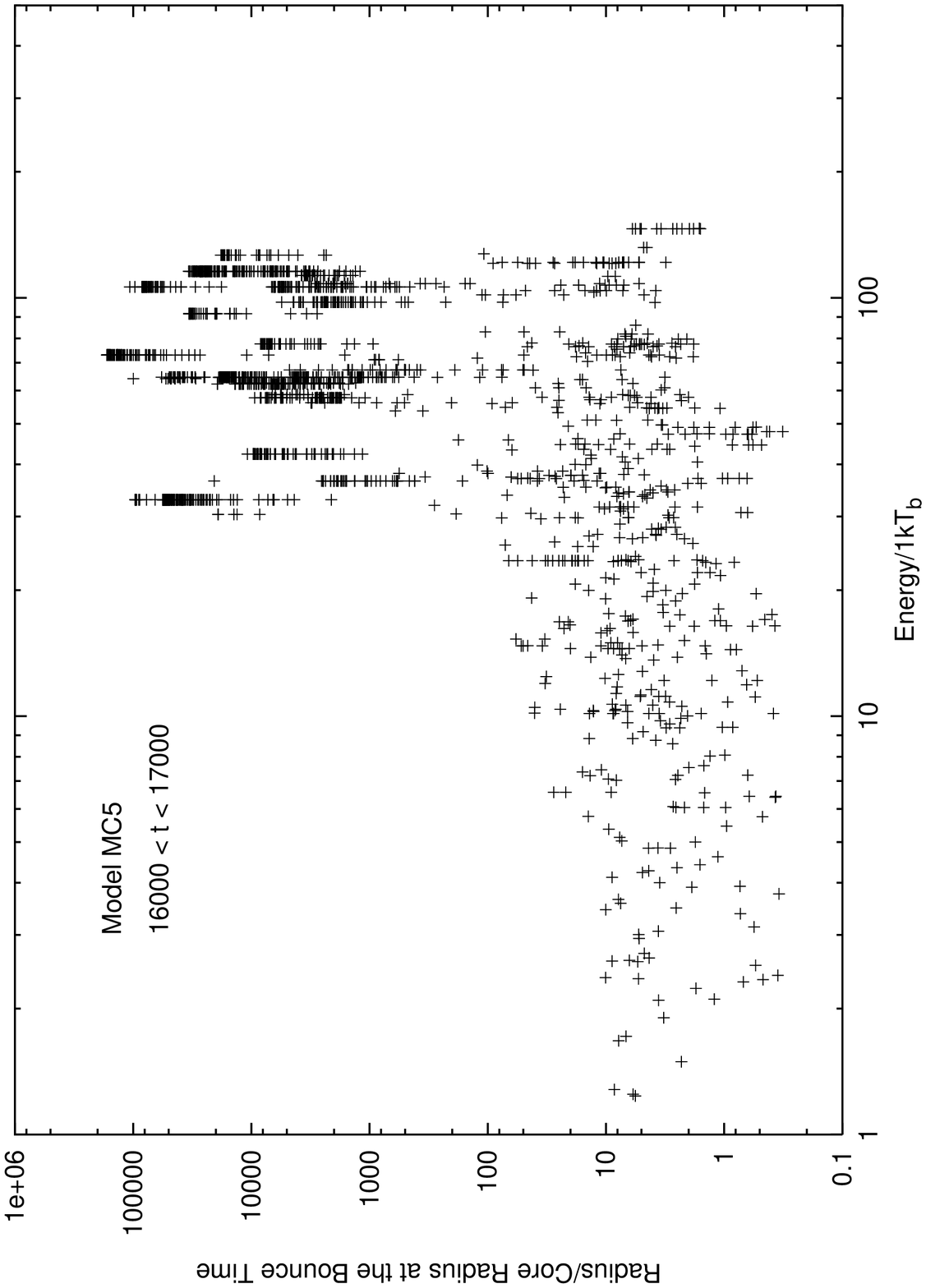}
\includegraphics{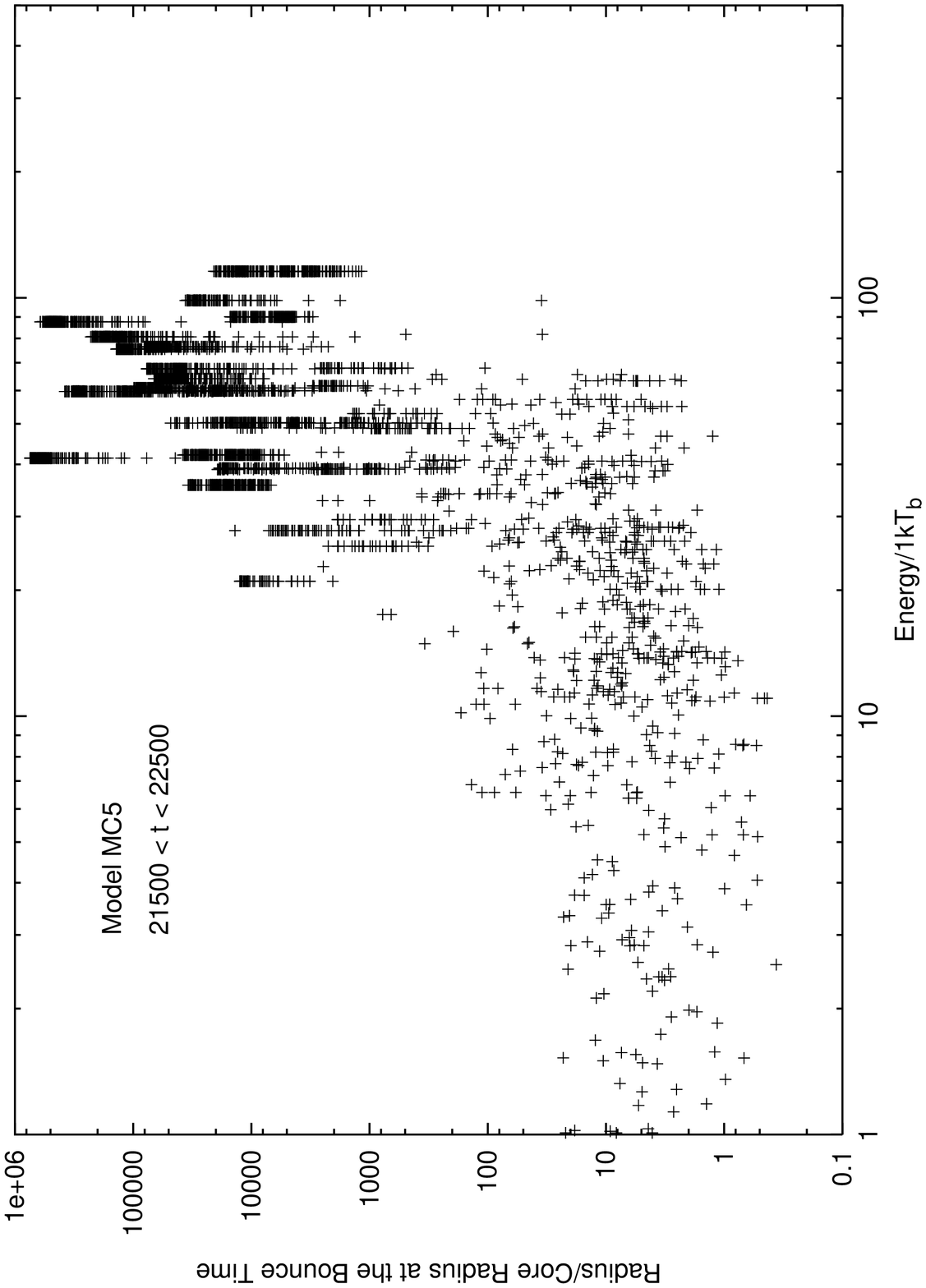}

 \caption{Model MC5, binary distribution in radius over
 core radius vs. binding energy in $kT$ plane, for four different
 times as indicated in the plots; note the distinct sequences occurring due to
 the oscillations of the central density.}
 \label{f28}
\end{figure*}

\begin{figure*}
\vskip 13truecm
\includegraphics{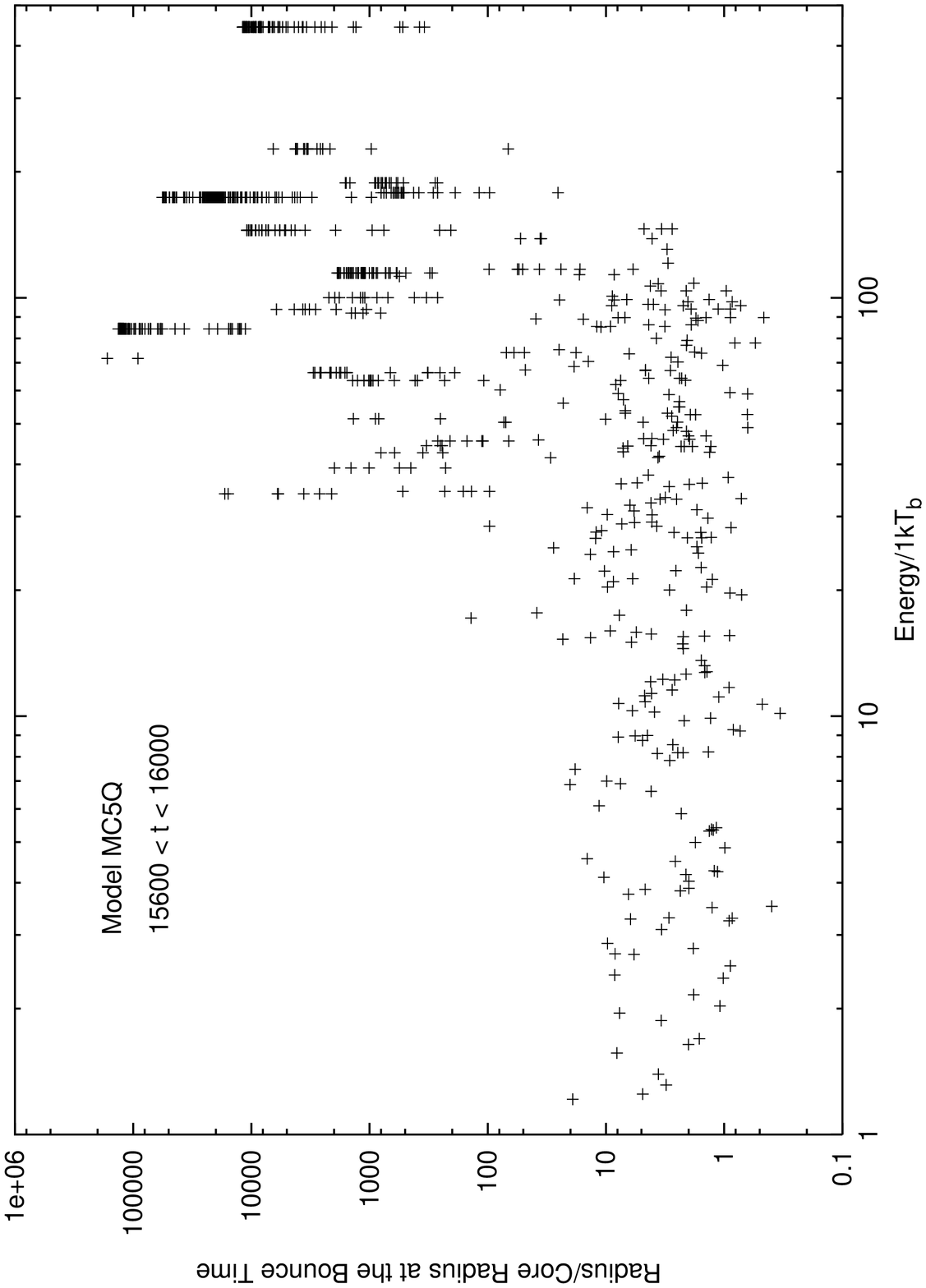}
\includegraphics{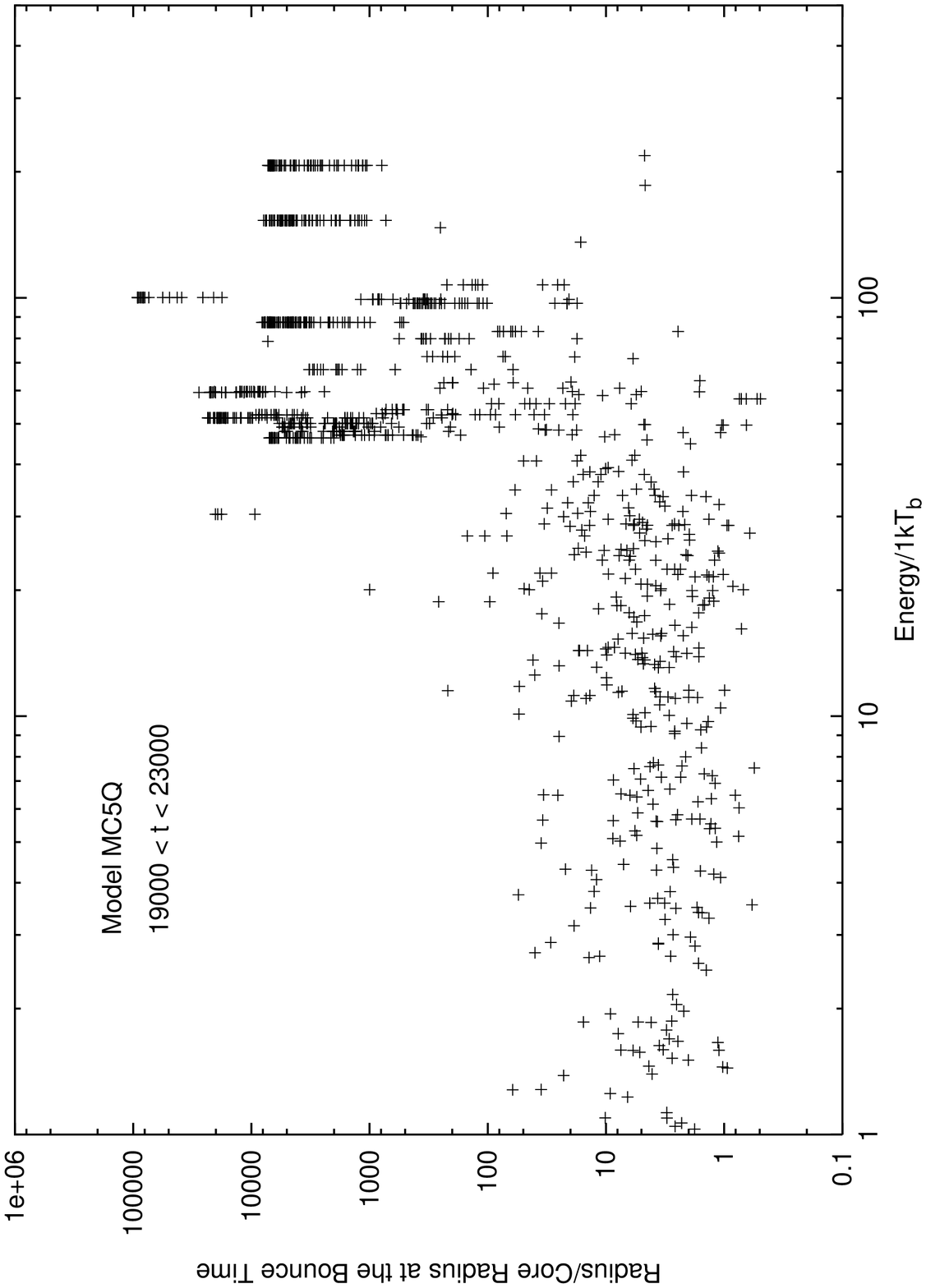}
\includegraphics{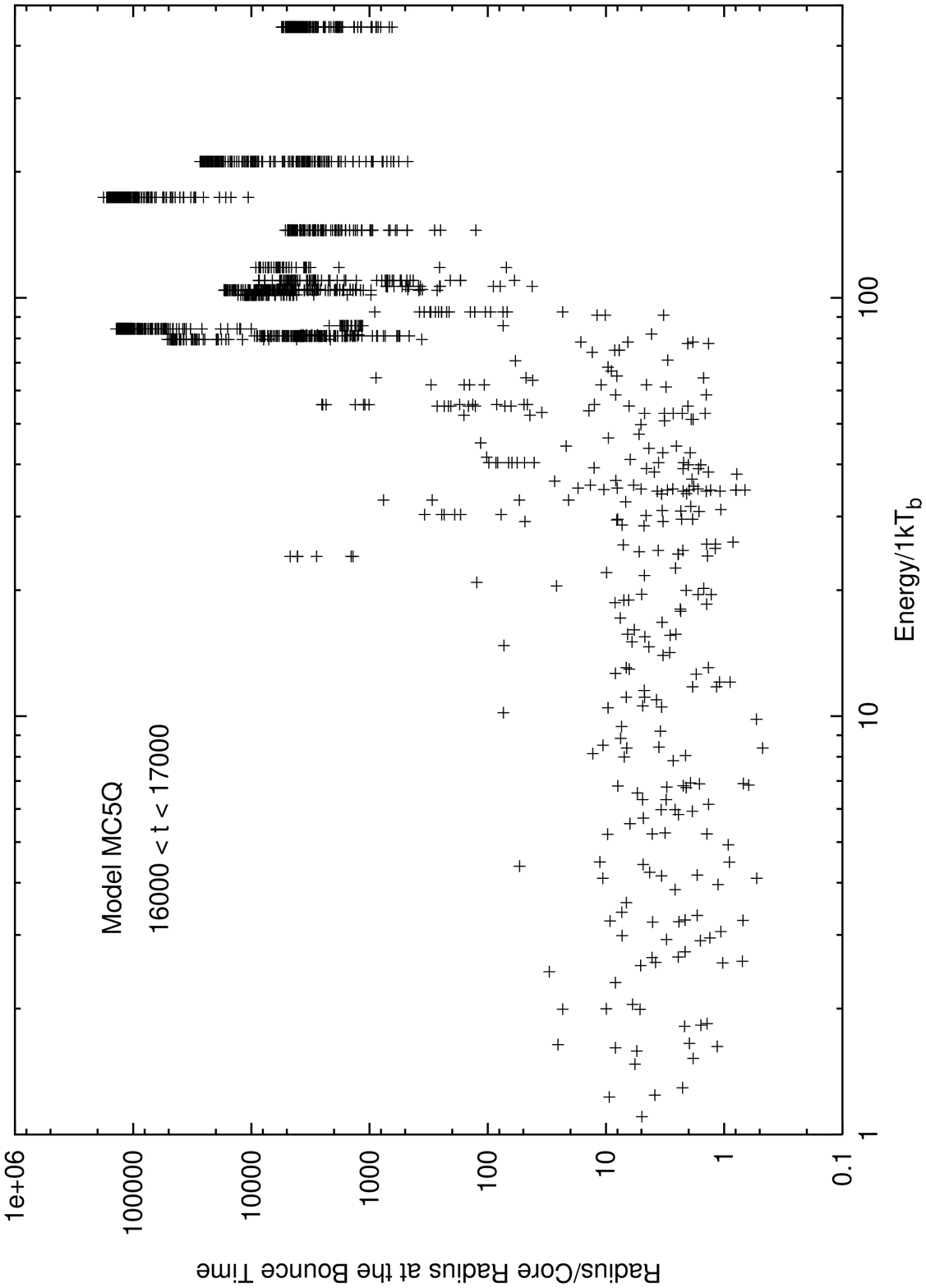}
\includegraphics{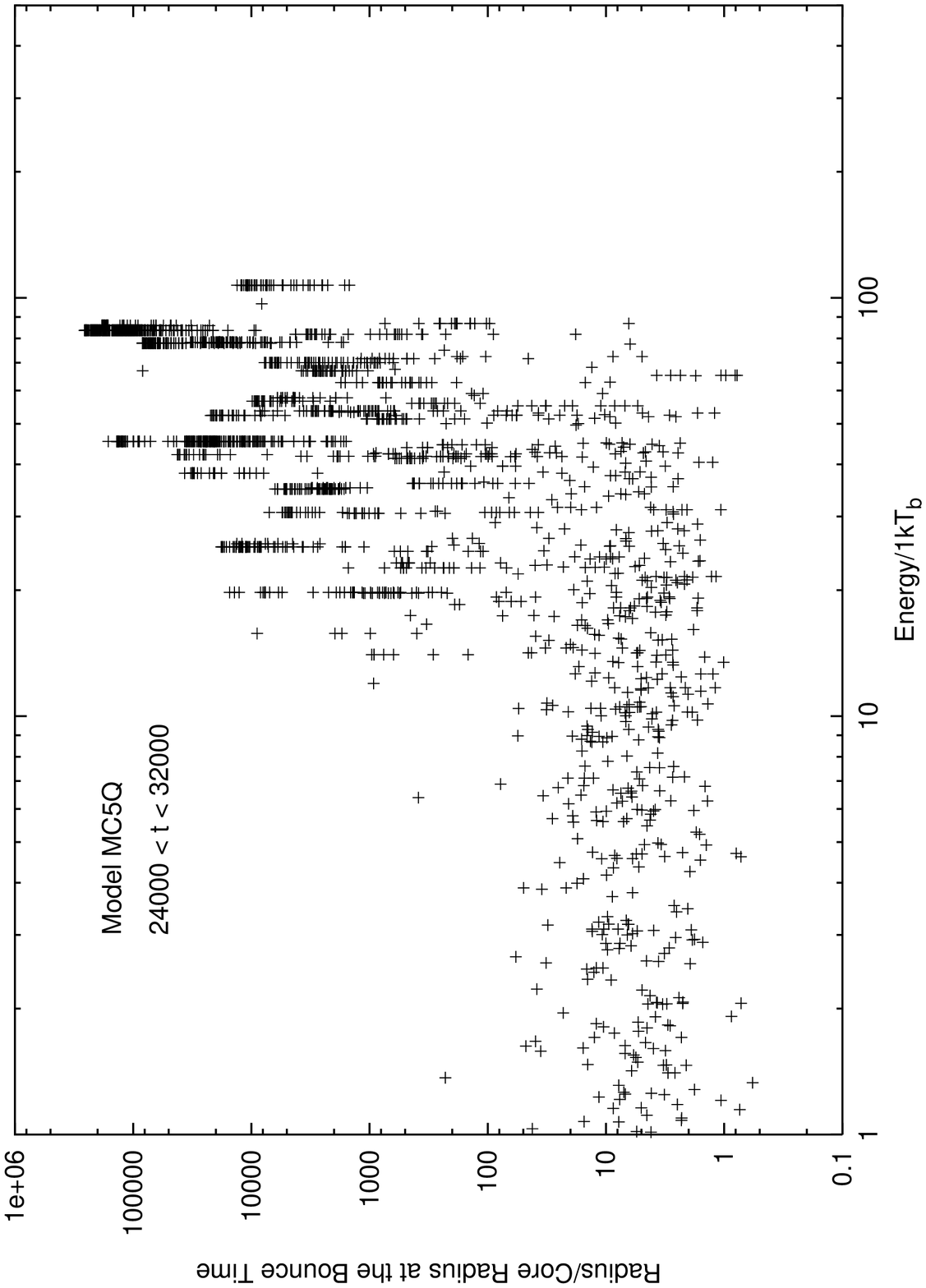}
  \caption{Model MC5Q, binary distribution in radius over
 core radius vs. binding energy in $kT$ plane, for four different
 indicated times; note the distinct sequences occurring due to
 the oscillations of the central density.}
 \label{f32}
\end{figure*}

 In Figs. \ref{f21} and \ref{f22} crosses show the events of strong
 interactions  between binaries and field stars and other binaries,
 together with the time evolution of the central density. It is
 clear that most of these interactions take place during the phases of maximum
 density, consistent with the picture of gravothermal oscillations.
 Here our runs also agree with results concerning the
 oscillations observed by Takahashi \& Inagaki (1991) for a stochastic
 Fokker-Planck model (stochastic binary formation and energy generation of
 binaries, but no binary orbits and relaxation interactions with themselves
 and with single stars considered), by
 Makino (1996) for $N$-body simulations and by Giersz (1998) for full
 Monte Carlo
 simulations. Note that maxima of the central density observed in our
 model are a few orders of magnitude larger than in the full Monte Carlo model.
 This is connected with the way in which the central density is computed in the
 latter. It is estimated from the position and masses of a few
 innermost stars (for details see Giersz 1998), which can lead to an
 underestimation
 of the density. On the other hand, the high 
 spatial resolution of the gaseous model in the core (where the innermost shell
 could consist of a mass of only a small fraction of a single star),
 and its extremely short time step during maximum collapse,
 both lead to very high central densities. It is worth mentioning
 that the refrigeration cycles (characteristic feature of gravothermal
 oscillations for continuum models already shown in Bettwieser \& Sugimoto
 1984) are also present in our models. Though they are
 very noisy, at least four distinct cycles can be separated, which correspond
 to four different oscillation periods. Such periods occur from period 
 doublings as they were studied when increasing $N$ in detailed
 Fokker-Planck and gaseous models of gravothermal oscillations 
 (Heggie \& Ramamani 1989,
 Breeden et al. 1994, Spurzem 1994). The shape and the size of the
 loops in the refrigeration cycle are consistent with those in the pure Monte
 Carlo runs, though the latter showed even more fluctuations. It is
 remarkable, that the stochastic events occurring in both Monte Carlo
 models do not disturb much the path of the entire system in phase space,
 as it is provided by the continuum models.

 In Figs. \ref{f28} for model MC5 and Figs. \ref{f32}
 for model MC5Q we provide snapshots of the binary distributions
 in energy-radius space of the cluster for four different time sets (see labels
 in the figures). The crosses shown in the figures represent all binaries
 which have interactions during the stated times (the number of crosses depend
 on the state of the system - large density means more interactions, 
 small density
 less interactions). For model MC5 a bimodal distribution of binaries can be
 clearly
 seen - there are binaries with high binding energy far from the core 
 and binaries in
 the vicinity of the core with a wide distribution of binding energies. In the
 course of evolution the build up of a reservoir of binaries
 in the outer halo (binaries in so--called parking orbits, as
 noted already by Hut, McMillan \& Romani 1992) is clearly visible,
 and will be even more pronounced in models with many primordial binaries
 presented here in subsection 3.4. Such
 binaries have orbits extending very far out into the halo,
 and do not enter the core
 (where strong interactions preferentially take place). Binaries on
 parking orbits can be seen in the snapshot as
 vertically aligned sequences of cross symbols
 at the same constant binary binding energy. This occurs because in the
 Monte Carlo procedure the binaries in a nearly invariant
 orbit are repeatedly picked in different positions.
 Newly formed binaries appear in the bottom left corner of the figures
 and then in the course of evolution (interactions with single stars) they are
 migrating in the direction to the right and upwards.
 Finally they are ejected in a single strong
 interaction or put in a parking orbit in a less strong interaction. Note
 the rather large number of binaries in parking orbits. In the full Monte
 Carlo models a much smaller number of such binaries is observed (only a
 few). The
 reason for this discrepancy is not clear, and may be connected with
 the already mentioned observation of higher central densities 
 (leading to larger binary formation probability) in the
 maxima for our model as compared to pure Monte Carlo and $N$-body models.
 For model MC5Q the
 build up of the reservoir is also clearly visible, but with smaller number
 of binaries in parking orbits. Because of the presence of
 4b encounters, which each destroy one binary,
 the number of binaries in the system drops faster, making the strip
 in the bottom of the figures less populated. Nevertheless, 
 the bimodal distribution
 of binaries (parking orbits and core and its vicinity) is still well visible.
 The presence of strong 4b interactions in the system creates a wider
 distribution of binaries in energy (note that in each 4b
 interaction, while one binary is destroyed, the other is considerably
 hardened) as can be seen for example from
 comparison of the upper right plot of Fig. \ref{f28} 
 and Fig.~\ref{f32} for model MC5 and MC5Q,

 respectively.

 In the next two subsections we will discuss the Monte Carlo stochastic models
 with many 
 primordial binaries and compare their results with data available in the
 literature (HA92, GGCM91).

 \subsection{Heggie's Runs}

\begin{figure}
\psfig{figure=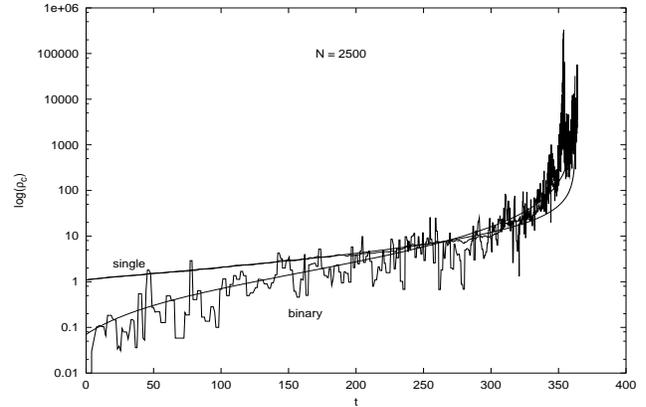,height=5.5cm,width=8.5cm,angle=-90}
 \caption{Comparison of two-component gaseous model
 with stochastic Monte Carlo binary model; 
 central densities of single stars and
 binaries as a function of time, fluctuating curves belong to stochastic model,
 see further explanation in the text.}
\label{f1}
\end{figure}

 \begin{figure}
 \psfig{figure=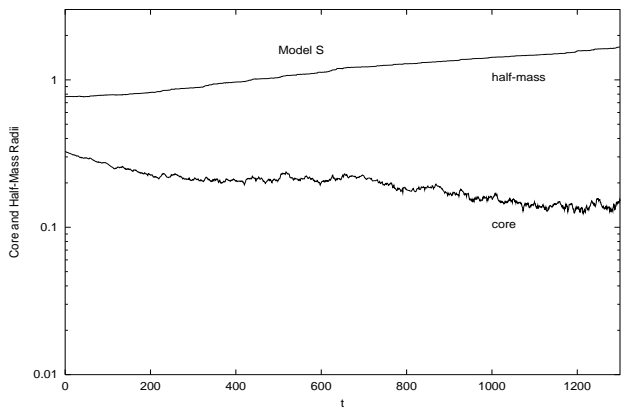,height=5.5cm,width=8.5cm,angle=-90}
 \caption{Model S, core and half-mass radius of the single
 stars as a function of time.}
 \label{f2}
 \end{figure}

 \begin{figure}
 \psfig{figure=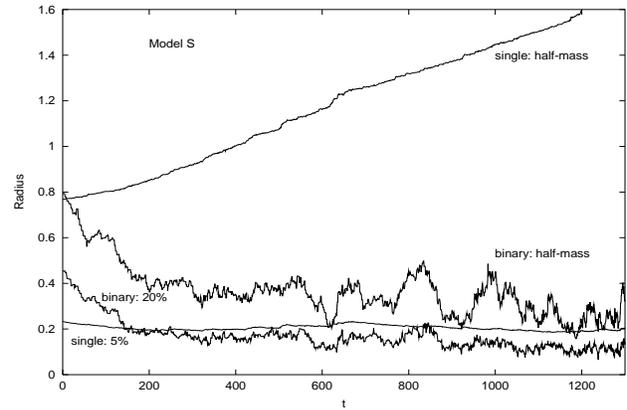,height=5.5cm,width=8.5cm,angle=-90}
 \caption{Model S, Lagrangian radii for the single stars 
 and binaries
 (radii containing the innermost 5\%, 50\% of the single stars, and
 20\%, 50\% of the bound binaries remaining in the cluster) as a function of time.}
 \label{f3}
 \end{figure}

 \begin{figure}
 \psfig{figure=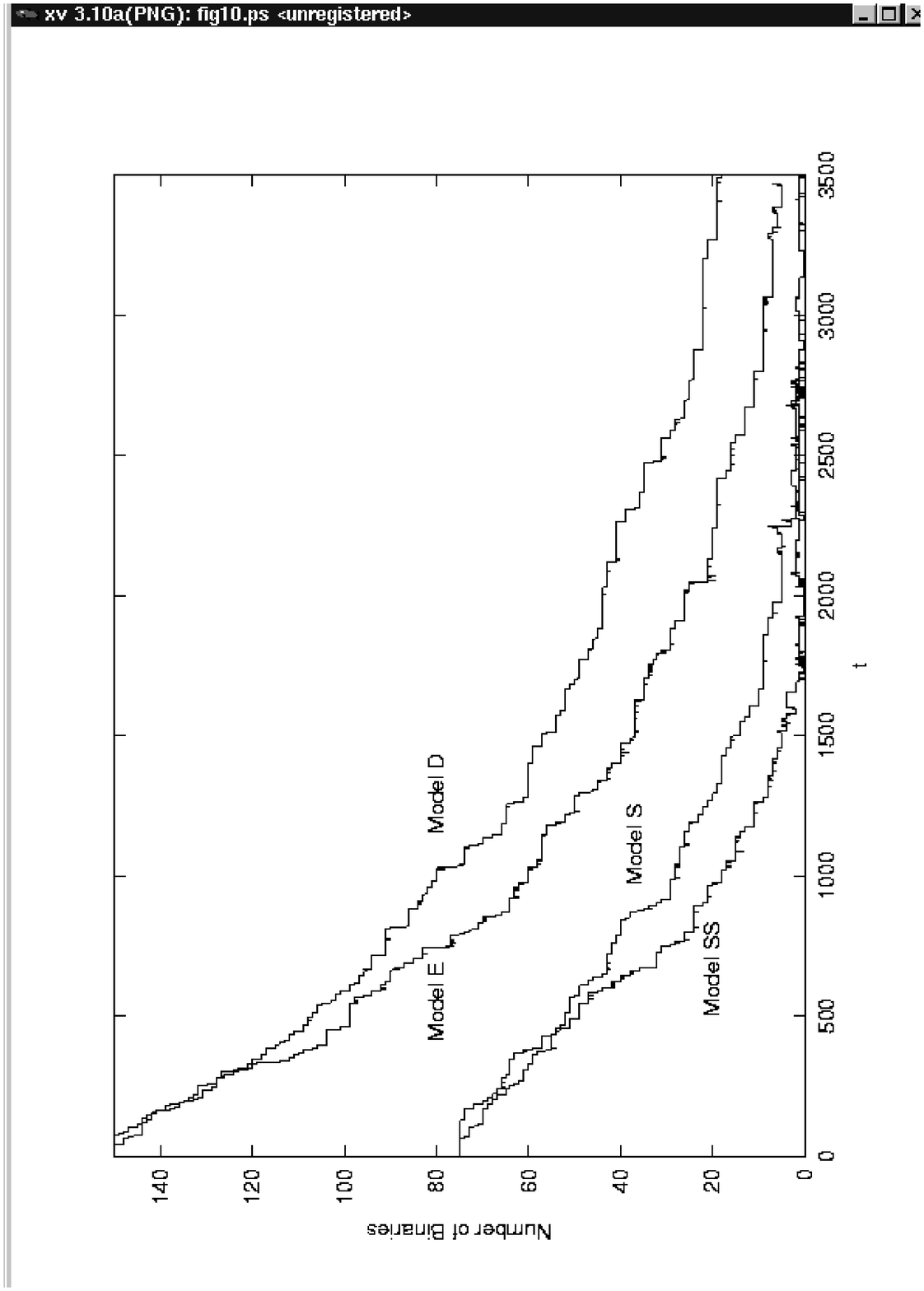,height=5.5cm,width=8.5cm,angle=-90}
 \caption{Model S, D, E, and SS; number of bound binaries as a 
 function of time.}
 \label{f4}
 \end{figure}

 \begin{figure}
 \psfig{figure=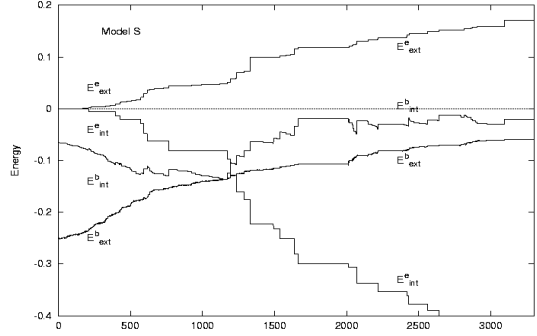,height=5.5cm,width=8.5cm,angle=-90}
 \caption{Model S, energy balances as a function of time 
 in $N$-body units. 
 The four different contributions are described in the text.}
 \label{f5}
 \end{figure}
 
 \begin{figure}
 \psfig{figure=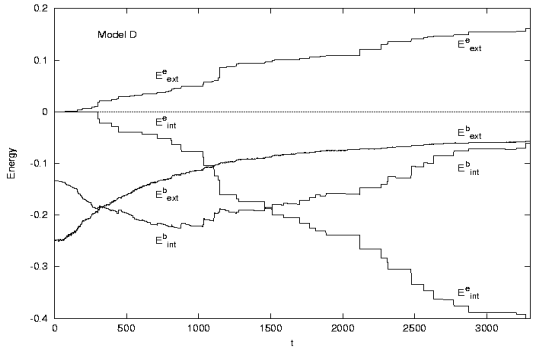,height=5.5cm,width=8.5cm,angle=-90}
 \caption{Model D, energy balances as a function of time, 
 as in Fig. \ref{f5}}
 \label{f6}
 \end{figure}

 \begin{figure}
 \psfig{figure=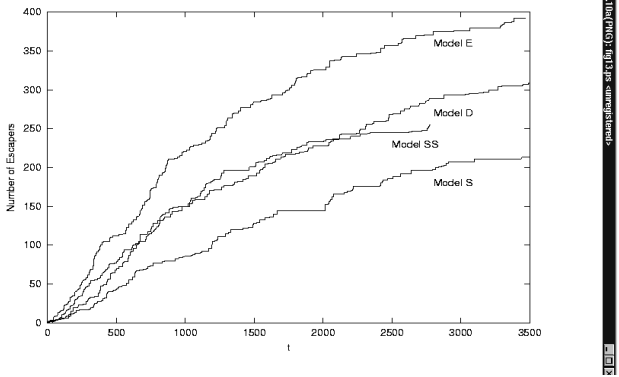,height=5.5cm,width=8.5cm,angle=-90}
 \caption{Model S, D, E, and SS; total number of escaping stars
 in singles and binaries as a function of time.}
 \label{f7}
 \end{figure}
 \begin{figure}
 \psfig{figure=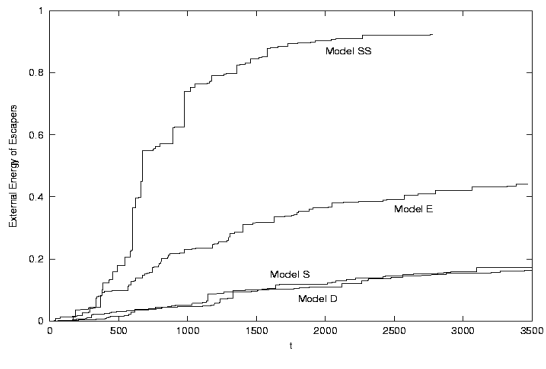,height=5.5cm,width=8.5cm,angle=-90}
 \caption{Model S, D, E, and SS; total external
 energy of escaping singles and binaries as a function of time.}
 \label{f8}
 \end{figure}

 \begin{figure}
 \psfig{figure=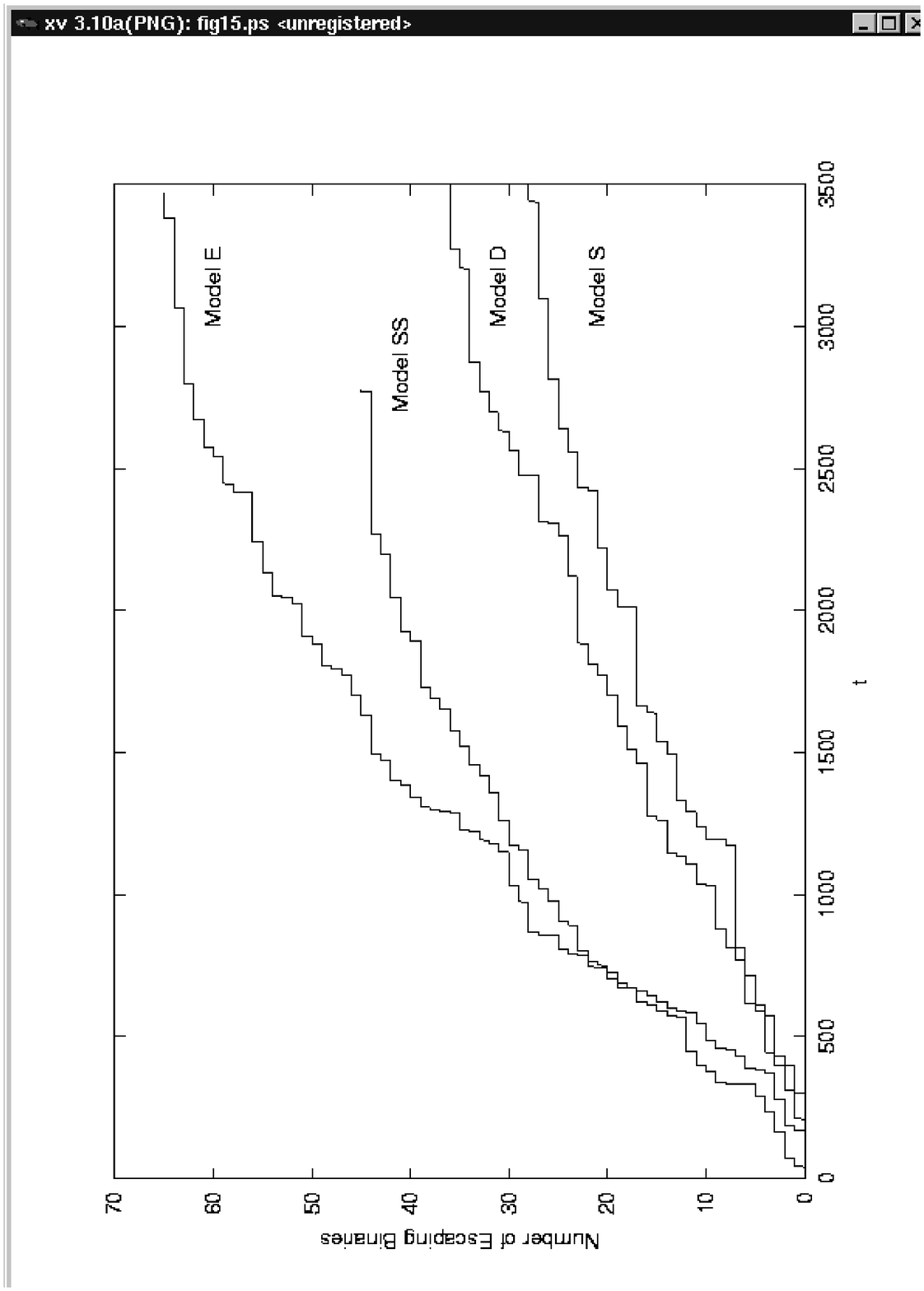,height=5.5cm,width=8.5cm,angle=-90}
 \caption{Model S, D, E, and SS;  number of escaping binaries
 as a function of time.}
 \label{f9}
 \end{figure}

 First, we have done one more test model in addition to those
 already discussed in Paper I, checking the correct mass segregation
 rate of a two-component system consisting of stars of masses
 $m_1=1$ and $m_2=2$. Hence the mass ratio is two, which
 mimics the situation of Heggie's model (HA92, see following Sect.)
 for which single stars and binaries are all composed of stars
 of the same mass. In Fig. \ref{f1} we show the central density evolution as
 a function of time for a two-component gaseous model (see Spurzem \&
 Takahashi 95, Spurzem 1992, the treatment is also equivalent to the same
 kind of gaseous model test runs presented in HA92) and for
 stochastic Monte Carlo model
 (using a gaseous model only for the single star component).
 To isolate the mass segregation effects only, all close
 3b and 4b encounters were artificially
 suppressed in our model. As one can see there is an excellent agreement
 between both models over many orders of magnitude in
 the central binary density. The fluctuations
 of the binary density in the stochastic model are not present in the
 continuous
 gaseous model. There is also a few percent difference in the final
 core collapse time. It is of the same order as usual
 differences between collapse times of either different physical models
 (Fokker-Planck, $N$-body, gas), or even the scatter of collapse
 times between stochastic models themselves starting with different
 realizations of the initial model (using other random number sequence,
 see a detailed discussion of this effect in GHI).

 Now we discuss a number of simulations which may be considered as test
 cases for our stochastic binary treatment in the refined version of
 the stochastic Monte Carlo
 method (as compared to Paper I). We denote these models
 by S, D, E, and SS, with the same naming as given in
 Table 1 of HA92. They are runs with 2500 stars each,
 containing 75 primordial binaries with binding
 energies in the range 2 to 20 kT (logarithmically
 equally distributed), where the spatial distribution of the binaries
 is the same as that of the single stars (Plummer's model).
 While this describes the standard model S, there are model D
 (with double the number of binaries, but the same binding
 energy distribution), model E (double number of binaries,
 but stretched from 2 to 200 kT), and model SS (only 75 binaries
 as the standard model, but binding energy distribution extended
 from 2 to 2000 kT).

 \begin{figure}
 \psfig{figure=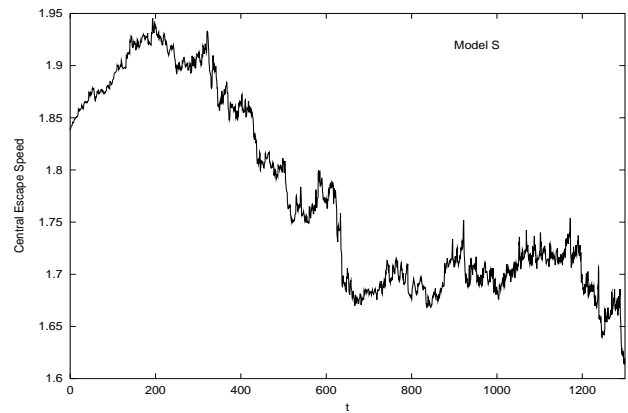,height=5.5cm,width=8.5cm,angle=-90}
 \caption{Model S, central escape speed as a function of time.}
 \label{f10}
 \end{figure}

 \begin{figure}
 \psfig{figure=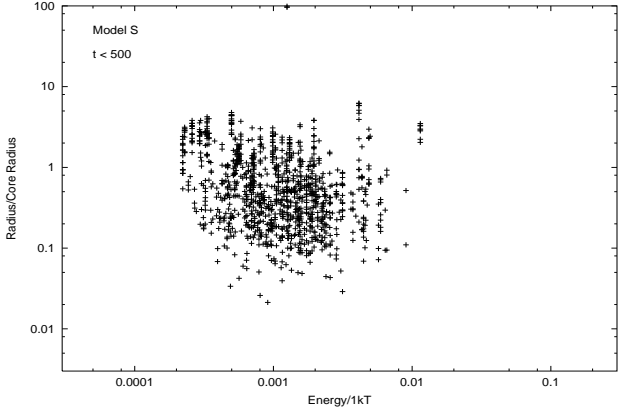,height=5.5cm,width=8.5cm,angle=-90}
 \caption{Model S, binary snapshots in a diagram plotting
 radius over initial core radius against binding energy in $kT$
 (initial), for the first
 500 time units. There are more points than actual binaries because data
 of several time outputs have been plotted together.}
 \label{f11}
 \end{figure}

 \begin{figure}
 \psfig{figure=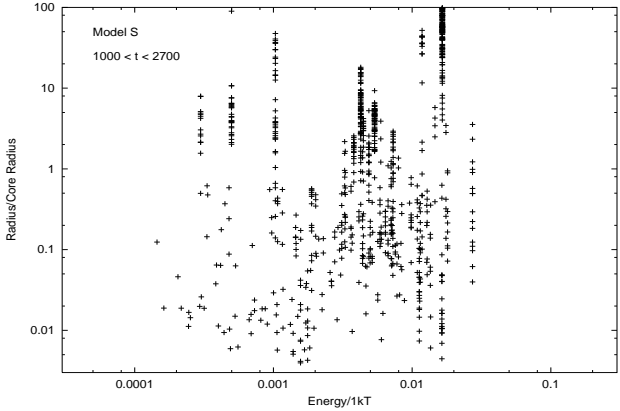,height=5.5cm,width=8.5cm,angle=-90}
 \caption{Model S, as Fig. \ref{f11}, but for time $1000<t<2700$.}
 \label{f12}
 \end{figure}

 \begin{figure}
\psfig{figure=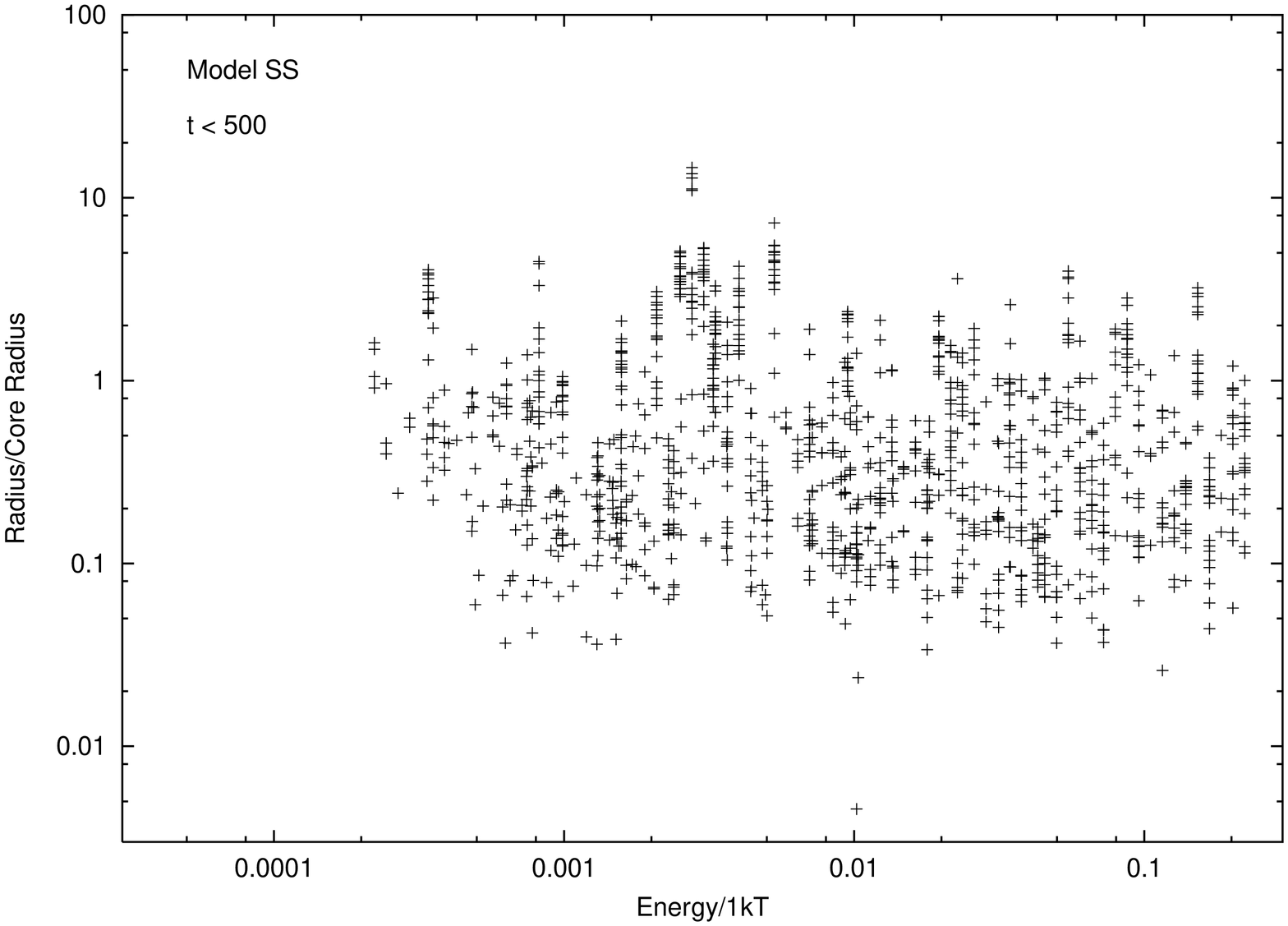,height=5.5cm,width=8.5cm,angle=-90}
 \caption{Model SS, as Fig. \ref{f11}}
 \label{f13}
 \end{figure}

 \begin{figure}
\psfig{figure=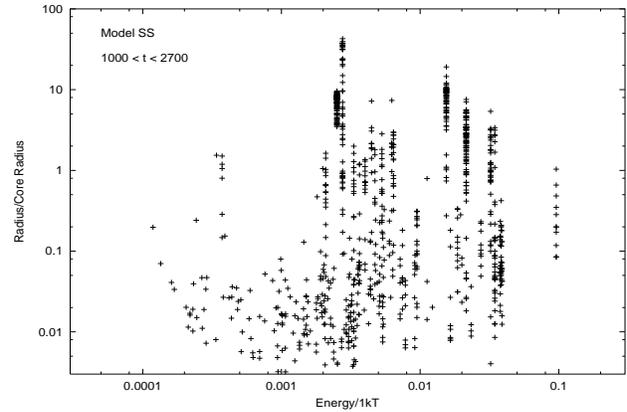,height=5.5cm,width=8.5cm,angle=-90}
 \caption{Model SS, as Fig. \ref{f12}}
 \label{f14}
 \end{figure}

 All our results match rather closely those of HA92,
 though each run takes only of the order of one hour on a Pentium II
 computer. We do not want to elaborate more on the physical interpretation,
 because this can be referred to HA92. Instead we present
 a number of figures in exactly the same manner as there, such that
 the reader can judge how well
 an agreement can be reached by our
 Monte Carlo type model. Clearly at such low particle numbers
 we cannot expect a complete match of the results. However, we
 have performed several runs for each model with varying random
 number seeds and our general conclusion is, that some features are
 stable while others exhibit stochastic variations. In our following
 arguments we will always try to discuss which differences to the
 $N$-body models we consider as ``real'' (in the sense that they
 may still point to deficiencies of our stochastic Monte Carlo
 model) and which we consider as a result of the statistical
 variations at small $N$.

 Fig. \ref{f2} shows core and half-mass radii of the single stars in our system
 as a function of time (to be compared with Fig.~23 of HA92). 
 It should be noted, that for a
 comparison with HA92 here and in all other plots related
 to the core radius, one has to take into account that HA92
 use a non-standard definition of the core radius. They use
 a density radius $r_\rho$, similar to the prescription given in
 Casertano \& Hut (1985), 
 because it is
 operationally well defined for $N$-body simulations.
 In contrast to this the
 standard core radius $r_c^2=9\sigma_c^2/(4\pi G\rho_c)$, which
 we use here is in their
 $N$-body models more difficult to determine
 ($r_c$ is also sometimes called the King radius,
 where $\sigma_c$, $\rho_c$ denote the central 1D velocity dispersion
 and central density, respectively).
 As is stated in HA92 for example a Plummer model
 yields $r_\rho = 0.77 r_c$; but during the system's evolution
 this ratio changes. Therefore the core radii measured by
 HA92 (in their Fig.~23) in the late stages have a systematic
 trend to smaller values than those from our models.

\begin{figure}
 \psfig{figure=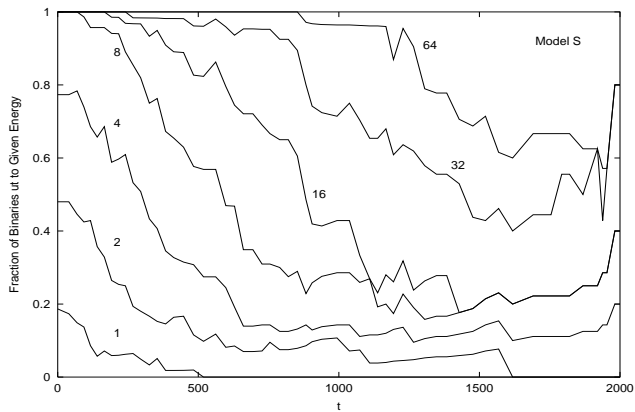,height=5.5cm,width=8.5cm,angle=-90}
 \caption{Model S, distribution of binding energies as 
 a function of time for all bound binaries. The unit of energy is 
 $2kT$, i.e. one and half times the initial 
 mean kinetic energy of single stars. Each line gives the fraction of these
 binaries with binding energy less than the stated value.}
 \label{f15}
 \end{figure}

 \begin{figure}
 \psfig{figure=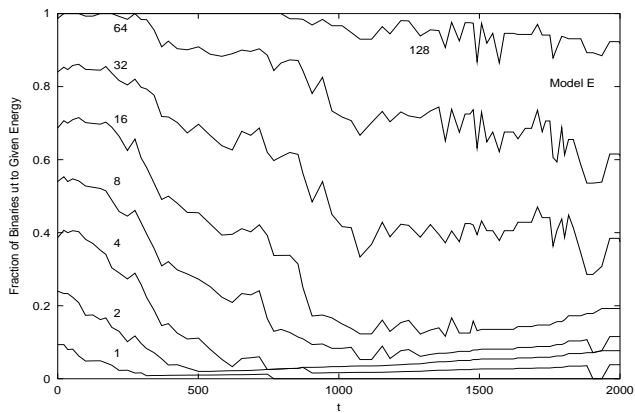,height=5.5cm,width=8.5cm,angle=-90}
 \caption{Distribution of binding energies of all binaries 
 in the  model E, see Fig. \ref{f15}}
 \label{f16}
 \end{figure}

 \begin{figure}
 \psfig{figure=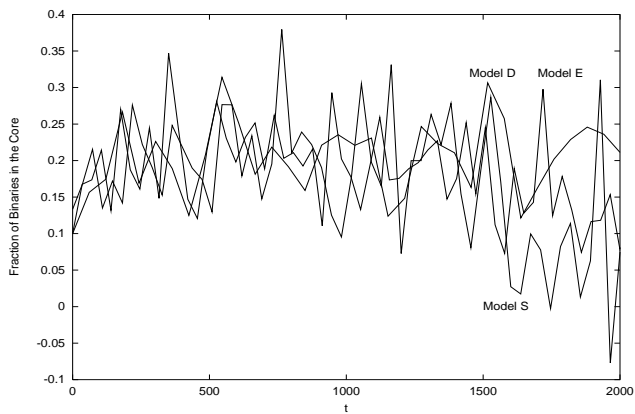,height=5.5cm,width=8.5cm,angle=-90}  
 \caption{Models S, D, and E; fraction of binaries in the core
 as a function of time.}
 \label{f17}
 \end{figure}

 \begin{figure}
 \psfig{figure=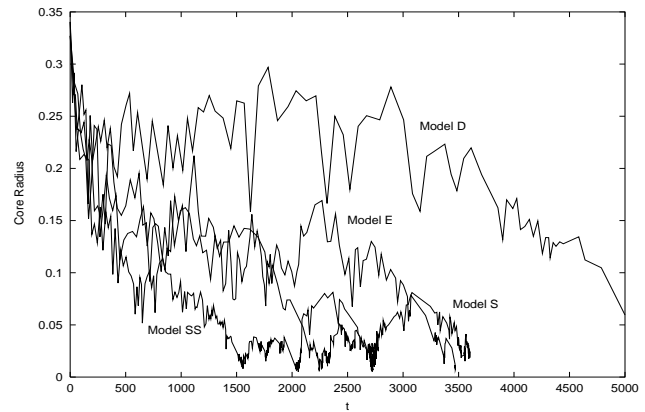,height=5.5cm,width=8.5cm,angle=-90}  
 \caption{Models S, D, E and SS; core radius as a 
 function of time.}
 \label{f18}
 \end{figure}

 Fig. \ref{f3} shows
 some Lagrangian radii separately for the single stars and binaries
 (see Fig.~8 of HA92). As in HA92 the segregation of binaries
 is very rapid up to about 100 time units. 
 A more stationary phase follows, characterized by
 continuous binary destruction, and energy generation, until
 most of the binaries are destroyed. For a more detailed physical
 discussion see HA92. The somewhat faster growth of the half-mass
 radius in the gaseous model (as compared to the $N$-body result)
 is a typical feature resulting from different outer boundary
 conditions (see GHI).

 Fig. \ref{f4} presents the number of binaries remaining
 bound in the system (see Fig.~9 of HA92). In contrast to
 HA92 after time $t\approx 1500$ (model E) or $t\approx 1000$
 (model S) the steady binary destruction does not slow down, but
 continues in our models to rather small binary numbers. In
 HA92 thereafter the binary number remains larger. The
 variation of such plots with 
 different random number initialization of the system has
 been explored. We
 found that it can explain to some extent
 the remaining differences, because
 stochastic variations are big in the late phases when there
 are only relatively few binaries left. The model we are using
 for our Fig. \ref{f4}, for example, underwent core bounce 
 relatively earlier than others, which explains the quicker
 destruction of binaries. Selecting another physically
 equivalent initial model can yield different collapse
 times, and very different features during the post-collapse phase
 (see GHI). We want to stress,
 however, that the global energy budget, presented for our model S
 and D in Figs. \ref{f5} and \ref{f6}, is in good
 agreement with the features seen in the direct $N$-body model
 (see Figs. 11 and 19 of HA92). For these plots we measure
 the total internal (binding) energy of the binaries remaining 
 bound in
 the system ($E^{\rm b}_{\rm int}$) and of those binaries
 which escaped ($E^{\rm e}_{\rm int}$), and the total
 external (i.e. potential plus translational) energy of
 all objects (singles and binaries) remaining bound
 in the system ($E^{\rm b}_{\rm ext}$) and escaping 
 ($E^{\rm e}_{\rm ext}$). Note that $E^{\rm b}_{\rm ext}$
 would be the standard total energy in a system
 without internal degrees of freedom (hence it starts
 at the canonical value of -0.25). As in HA92 we observe
 the initial phase of an increase of total energy due
 to hardening of the binaries in the system (until
 $t\approx 1000$ in Fig, \ref{f5}), which we would
 like to call the hardening phase, while afterwards
 total energy is gained in part by encounters leading
 to an escape of binaries with high binding energy. The
 total internal energy of bound binaries increases in this
 phase, which means that on average the typical binary
 in the system becomes less bound. We would like to
 call this the escaper phase. The reader is referred
 to a much more thorough discussion of the physical
 processes and balances determing these phases in 
 HA92. Here we just want to stress that our model is able
 to reproduce all features of the $N$-body model. 

 Now we get to Fig. \ref{f7}, where we have plotted the
 total number of escaping stars, where escaping binaries
 have been counted as two stars (see Fig.~12 of HA92). 
 Compared to our results 
 HA92 find much larger numbers, and no difference between
 model S and model D. They conclude that the presence of
 binaries does not enhance the rate of escape of single stars
 very much. In our isolated gaseous models we have not 
 yet incorporated single star escapers due to relaxation
 and tidal fields,
 which is a subject of future work.
 This means that differences in the escape
 rates between our four models S, SS, D, E should
 be interpreted only in relation to the close 3b and 4b
 encounter processes. Hence our model reveals
 differences regarding the origin of single star escapers,
 which cannot easily be seen in the complete 
 $N$-body of HA92, because their models are dominated by standard
 relaxation escapers. In contrast to the number of
 escapers, the external energy of escapers should in all
 models be dominated by the highly energetic 3b and 4b
 encounters. Consistent with this, our results for the
 external energy of all escaping objects (see 
 Fig.~\ref{f8} and Fig.~13 of HA92) agree
 much better with the HA92 model.
 The higher external energy of escapers for
 model E and SS can be attributed to the larger
 maximum binding energy of their binary energy
 distribution, which leads to very high recoil
 energies in 3b and 4b encounters.
 Also the number of escaping binaries, in our Fig. \ref{f9}
 matches well the HA92 results again (their Fig.~14).

 The central escape speed as a function of time,
 which is a direct measure of the central potential,
 is presented in Fig. \ref{f10}, as compared to Fig.~15
 of HA92. The $N$-body models of HA92 reach significantly
 deeper values of the central potential. We think that the
 reason for this, at least partly, is a
 bias originating from the pairwise potentials of the single stars.
 Hence the potential computed in the
 direct $N$-body model is more prone to clumps of particles,
 which we do not have in our gaseous model representation.
 Also the maximum is at a different time, in our model
 at the time of maximum binary segregation (roughly at $t\approx 200$)
 to the centre,
 in their model at the time of maximum collapse of the
 single stars ($t\approx 500$). 

 In Figs. \ref{f11} to \ref{f14} we provide snapshots of
 the binary distributions in energy-radius space of the
 cluster, for two different times, early just in the
 hardening phase, and late well into the escape phase, for
 model S and SS.
 In the late phases a bimodal binary distribution is observed,
 soft and hard binaries deep in the core, and a group of
 hard and very hard binaries being ejected into the very outer
 regions of the cluster. Very similar 
 behaviour can be seen in Fig. 22 of HA92, which gives the
 corresponding $N$-body data.

 Figs. \ref{f15} and \ref{f16} show the distribution of
 binding energies of the binaries as a function of
 time for models S and E 
 (see Figs.~16 and 20 of HA92); both results again agree rather well,
 except for the bins of the most bound
 binaries (between 32 and 128), which are more abundant in our model, due
 to the relatively early creation of new three-body binaries in
 the core. The time at which such first creation of
 new binaries occurs can vary strongly as a function of the
 random number initialization of the model.
 Note that the units
 of binary binding energy are different by an approximate factor of 2
 between HA92 and our models; corresponding lines can be identified
 from $t=0$.

 To complete our survey of results we present
 the fraction of binaries in the core (Fig. \ref{f17}) and
 the core radius (Fig. \ref{f18}) from the four models S,
 D, E, and SS (see in HA92 Figs.~17 and 18). 
 A quantitative comparison is
 difficult due to very noisy data of $N$-body and
 stochastic Monte Carlo models and due to the different and incompatible
 core definitions in both models. However, the initial increase
 of the binary fraction in the core, and subsequent slow decrease,
 are clearly visible in Fig. \ref{f17}. In Fig. \ref{f18} also
 the initial fast decrease of the core radii for all models
 is clearly seen, but the further evolution, particularly for
 model D is not fully compatible with results of HA92 (see their
 Fig.~18). Note that the above mentioned features, which are in good
 agreement, are related to the initial fast mass segregation phase
 of the binaries.

 \subsection{Gao's Runs}

 Gao's runs are named after the pioneering paper of Gao et al. (1991, GGCM91),
 which we use as a reference paper. This is the only paper providing
 models of live and self-consistent single and binary stars in a large
 number (30.000 primordial binaries) in a star cluster consisting
 of 360.000 stars (including the binary members). Their models were
 two-component Fokker-Planck models, treating the binaries as a smooth
 evolving mass distribution (since all stars have the same mass, only
 two-components are sufficient). The initial primordial binary
 distribution has a binding energy range from 3 $kT$ to
 400 $kT$, their spatial distribution is a Plummer model
 realization using the same scaling radius as for the
 single stars. In this subsection we describe, how
 we redo their calculation with our stochastic Monte Carlo model, what
 features can be reproduced, and which ones are different. We include
 the same physics as in GGCM91 as far as possible and reasonable. This
 means for example we use exactly the same 4b encounter cross-sections
 that they did. However, our model provides more details of the binary
 evolution and of the close encounters. For example, about the single
 star reaction products, which they could not provide due to the
 statistical treatment of their binaries.

 \begin{figure}
 \psfig{figure=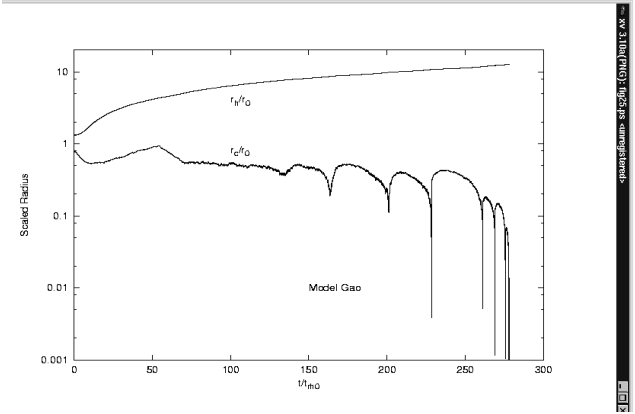,height=5.5cm,width=8.5cm,angle=-90}
 \caption{Evolution of the core and half-mass radii (scaled by
 the scale length of the Plummer model) as a function of time in
 units of initial half-mass relaxation time; this time unit is
 used in all following figures.}
 \label{f36}
 \end{figure}

 \begin{figure}
 \psfig{figure=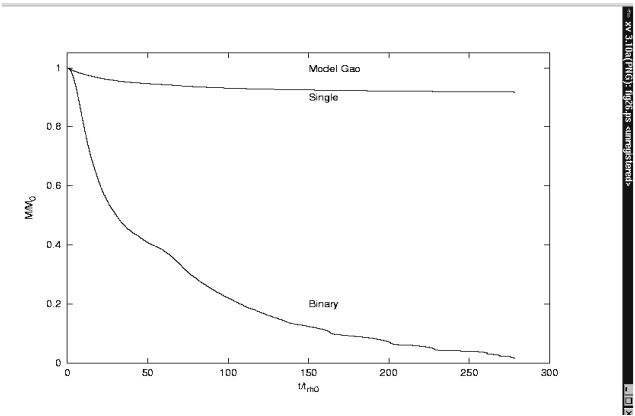,height=5.5cm,width=8.5cm,angle=-90}
 \caption{Mass fraction remaining in single stars and binaries as a
 function of time.}
 \label{f37}
 \end{figure}

 \begin{figure}
 \psfig{figure=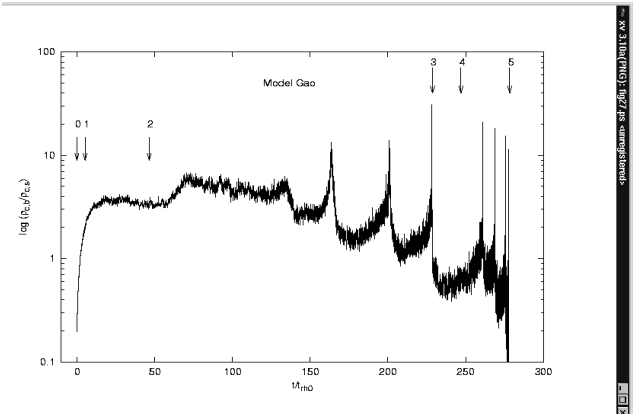,height=5.5cm,width=8.5cm,angle=-90}
 \caption{Ratio of central densities in binaries and singles as a
 function of time. Arrows indicate the times for snapshot plots of 2D and
 3D distributions of binaries bound to the system.
 The selection of times is explained in the text.}
 \label{f38}
 \end{figure}

 \begin{figure}
 \psfig{figure=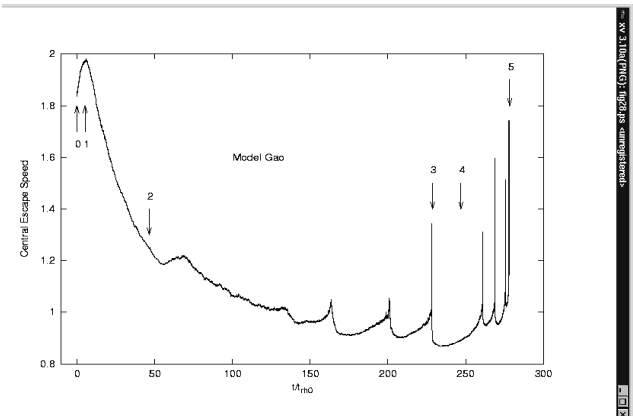,height=5.5cm,width=8.5cm,angle=-90}
 \caption{Central escape speed as a  function of time. Arrows as
 explained in the Fig. \ref{f38}.}
 \label{f39}
 \end{figure}

 \begin{figure}
 \psfig{figure=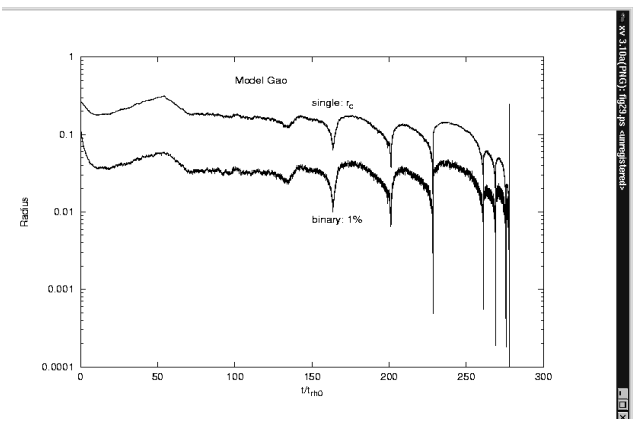,height=5.5cm,width=8.5cm,angle=-90}
 \caption{Evolution of the core radius of single stars and $1\%$
 Lagrangian radius of binaries as a function of time.}
 \label{f40}
 \end{figure}

 There is one technical remark to be made, regarding the very important
 choice of a proper relaxation interval for the Monte Carlo binaries
 (compare standard description in Paper I). 
 In contrast
 to Heggie's runs, here phases occur where there are very high densities
 of single stars and binaries. Binaries may stay in very elongated
 orbits, so if we pick binaries only at some fraction of their orbital
 time they are being ``left behind'' in an unphysical way in collapse
 phases of the single stars. In such a situation
 the results did not agree
 with other models or physical expectations. For those binaries newly
 obtained positions were very far outside, and practically no evolution
 of the binary population took place. Therefore we employ (after
 some experiments) the
 prescription to use the local relaxation time at two times the
 core radius, divided by ten, as the proper interval for relaxation.
 It would not be harmful to use an even smaller interval, but then
 more and more computational time would be spent to relax the binaries.
 A too small relaxation interval, however, would get into conflict with
 the Monte Carlo method's principles, because the picking of the
 binaries position at any point of the orbit becomes unphysical if
 done in a time interval much smaller than the orbital time scale.
 
 Fig. \ref{f36} shows the time evolution of the scaled core and half-mass
 radii of the single stars, to be compared with Fig.~1 of GGCM91. Like
 in their model the evolution can be divided into a first, rather fast,
 mass segregation phase of the binaries (as in the case of Heggie's models),
 a binary burning phase, in which gradually most of the binaries are
 destroyed or ejected by close 3b and 4b encounters, and finally a
 standard phase of gravothermal oscillations follows,
 which is dominated by the single stars only with the formation of a few
 new binaries due to 3b encounters. In this phase there are only
 a few hundred primordial binaries left in the 
 outskirts of the cluster, as will become clear later.
 The model continues until about 280 initial half-mass relaxation times
 (\tref). To assess what a heroic effort such a model would be
 in the direct $N$-body approach just note, that this corresponds
 to 0.8 million $N$-body time units or some 260.000 initial half-mass
 crossing times. We stop the model at about 280 \tref, because nearly 
 all binaries have been destroyed or ejected from the core, and the
 simulation goes on as a single star gravothermally oscillating model,
 where only binaries are formed in the high-density phases in the
 core, as in the Monte Carlo runs discussed in subsection 3.2.
 Different to those models, we only have a certain cloud of a few
 hundred binaries in parking orbits very far in the halo, but they
 do not influence the core evolution. Gao's runs reached a similar
 state already after 90 \tref; but note that the
 fraction of time spent in the different phases (such as
 oscillations or binary burning) seems at a first glance not
 much different from Gao's case. We will come back to this time scale
 problem below.

 First, we would like to refer the
 reader to the following figures, which show the mass fraction remaining
 in singles and binaries bound to the system (Fig. \ref{f37}), the ratio
 of central densities of binaries over single stars (Fig. \ref{f38}),
 the central escape speed as a measure of central potential (Fig. \ref{f39}),
 and the evolution of the core radius of single stars and the 1\%
 Lagrangian radius of the binaries (Fig. \ref{f40}), all as a function
 of time in units of \tref. Corresponding figures in GGCM91 are
 Figs.~2 (mass fractions), and 3 (ratio of central densities). We will
 now deduce the scenario of the evolution of our system from these
 figures, with additional information from pictures of the
 detailed binary distribution in energy (over initial $kT$) and radius
 (over initial core radius), provided for key time points
 in Figs. \ref{f44} and \ref{f45}. The time points are marked by
 arrows with numbers 0 to 5 in Figs.~\ref{f38} and
 \ref{f39}, corresponding times in \tref are given in the
 caption of Fig.~\ref{f44}. We will denote the binary snapshots
 at these times in short with snapshot number 0 to 5;
 they show the frequency of binaries
 (as a 3D surface, with contour levels projected onto the
 plane below) over a logarithmically equally spaced mesh in scaled
 energy and position (Fig.~\ref{f44}), or a direct projection of these
 data (one cross per individual binary) in Fig.~\ref{f45}.
 We deduce the following basic scenario:

\begin{figure*}
\vskip 17truecm
\includegraphics{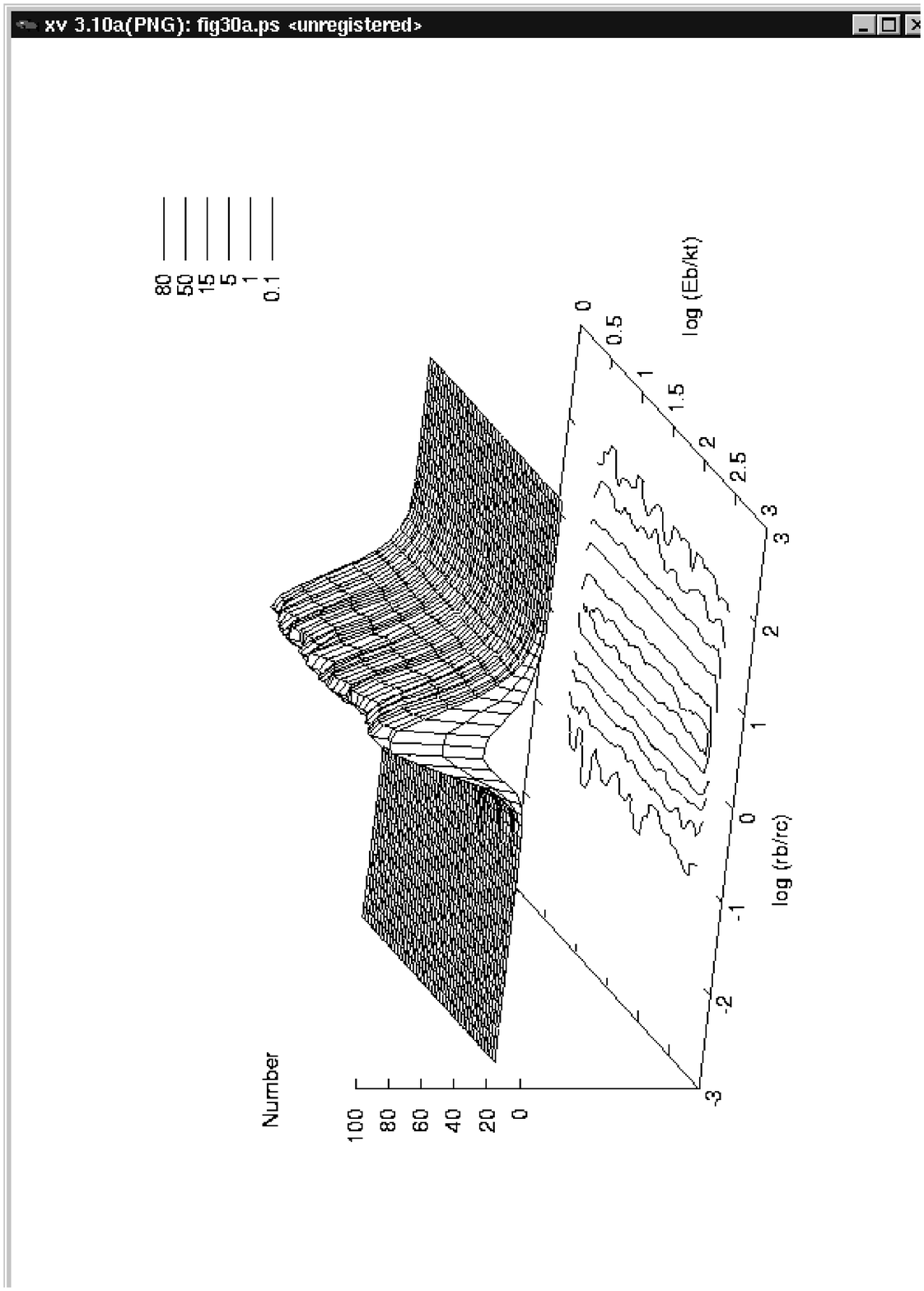}
\includegraphics{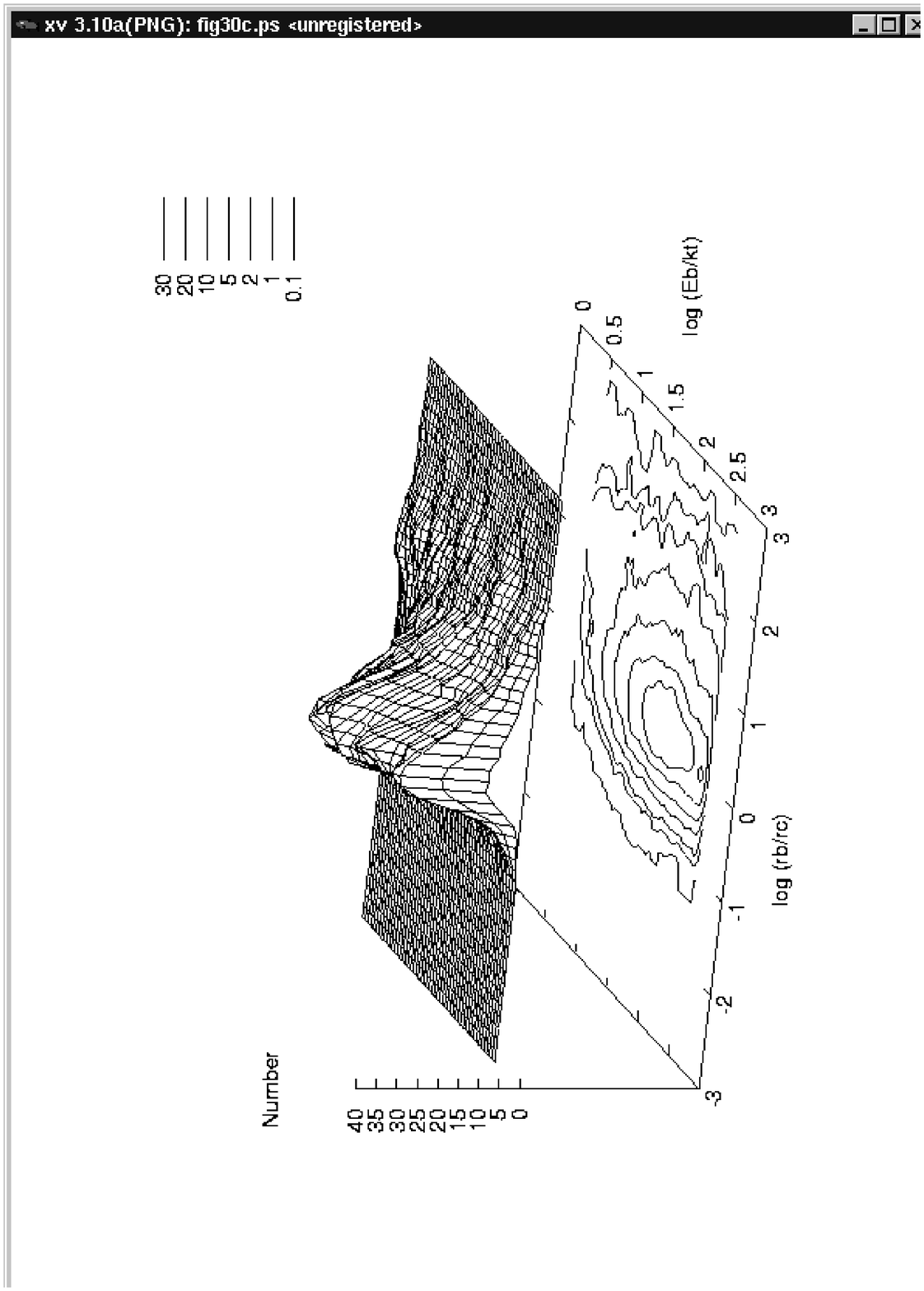}
\includegraphics{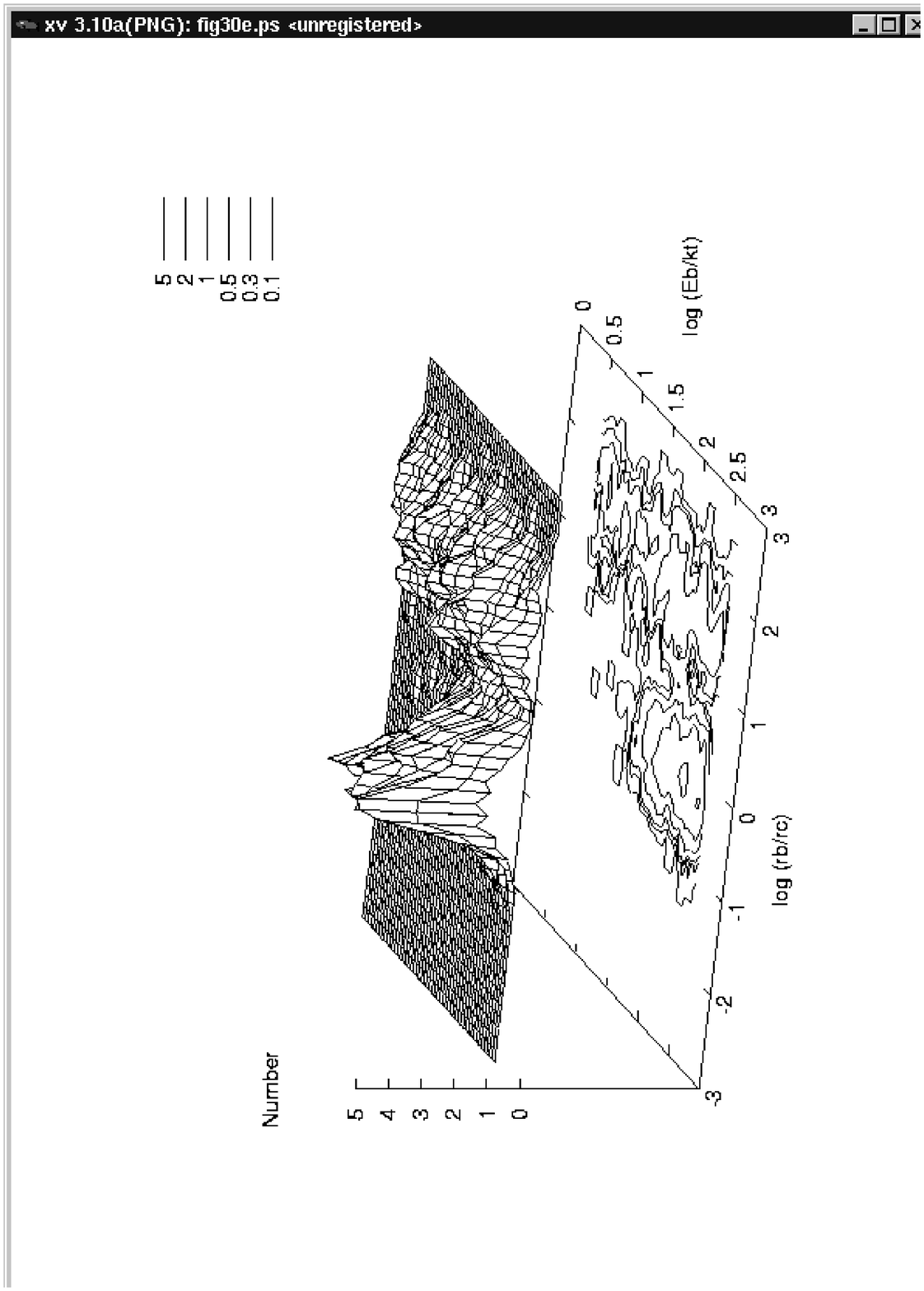}
\includegraphics{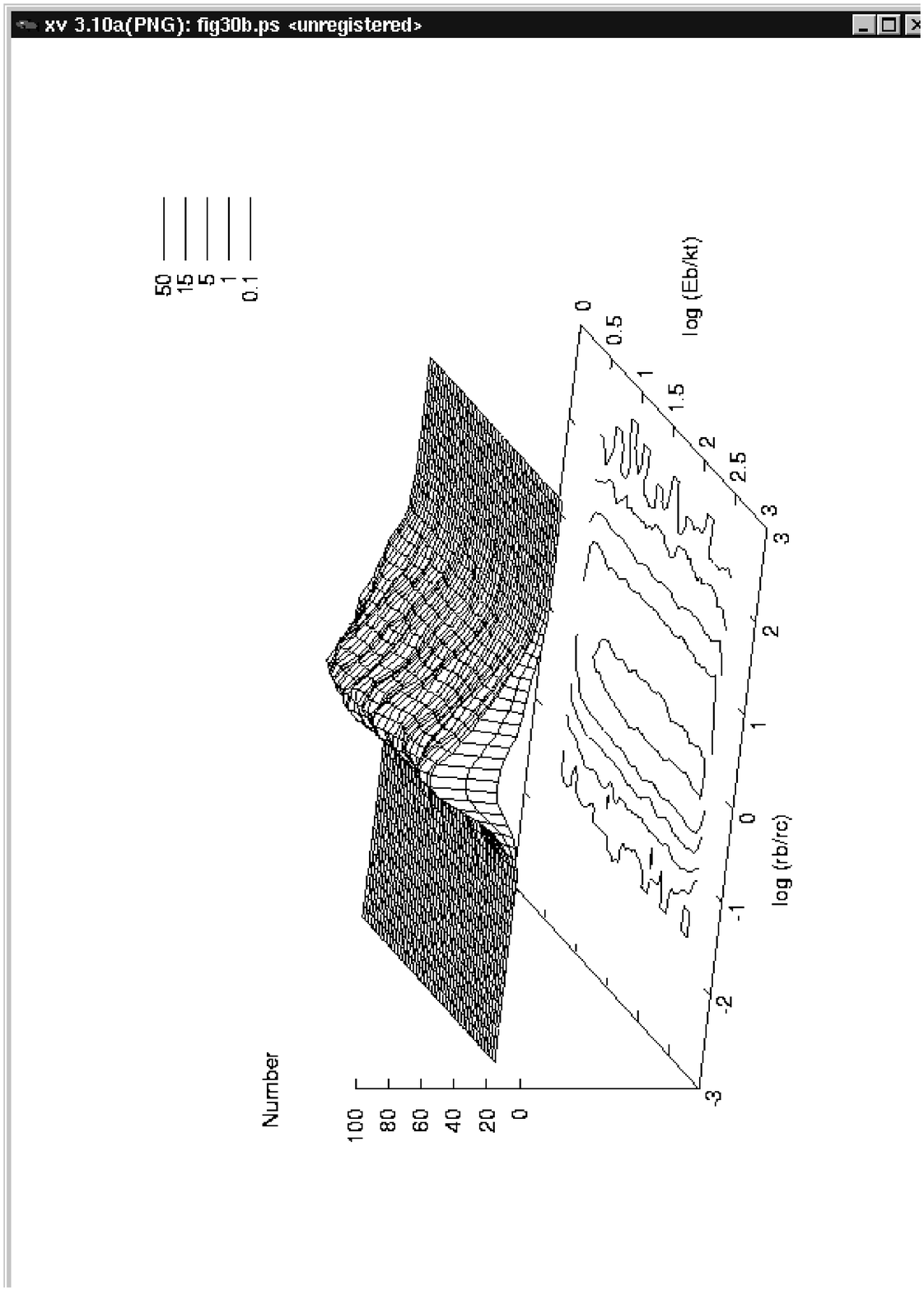}
\includegraphics{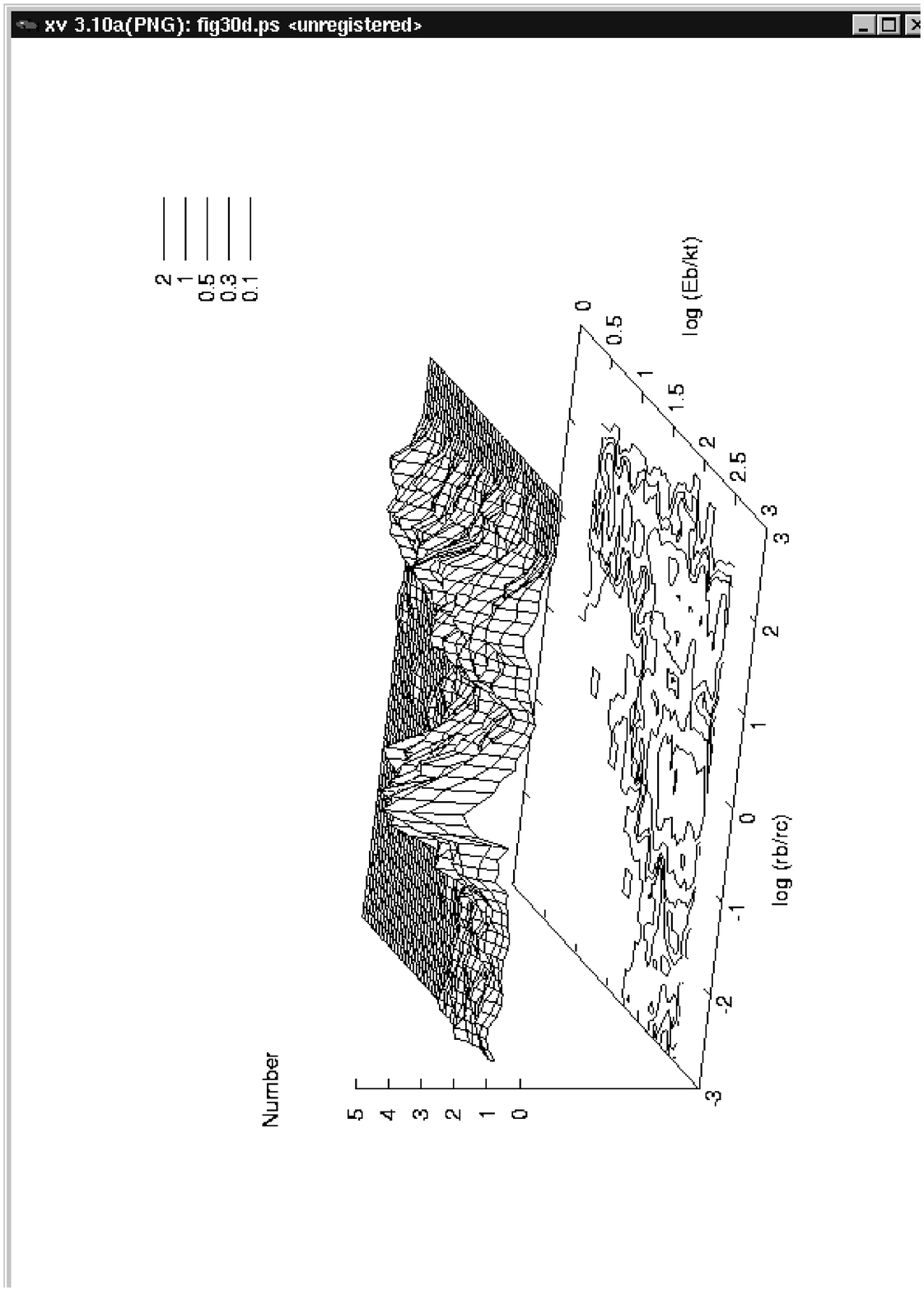}
\includegraphics{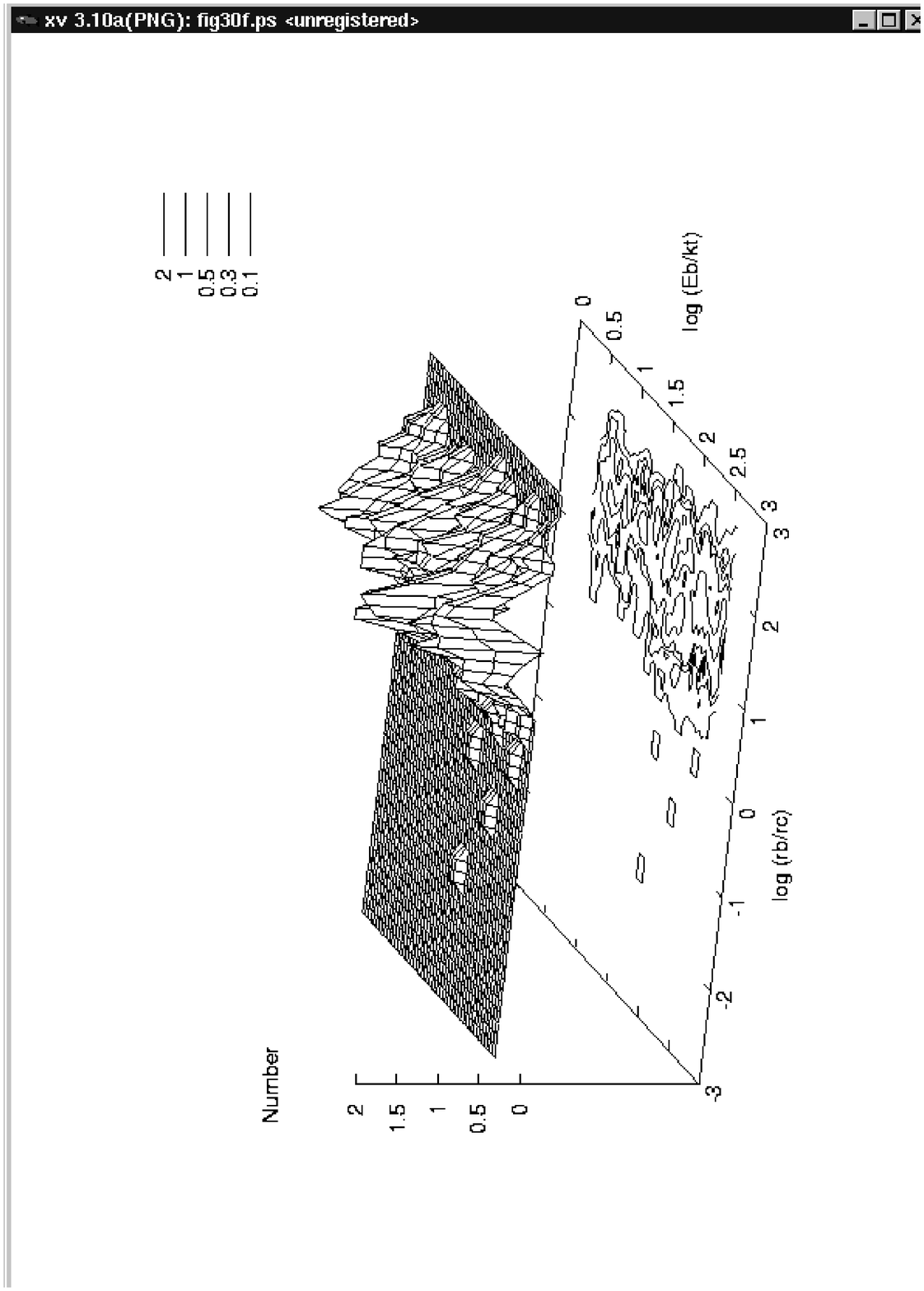}

\caption{Snapshot plots of 3D distributions of bound binaries. Plotted is
 the frequency of binaries in logarithmically equidistant bins of binding
 energy in $kT$ (initial) and position in units of initial core
 radius, for six different times in units of the
 initial half-mass relaxation time; top left 0, top right 5.21,
 middle left 46.51, middle right 228.70, bottom left 246.99, bottom
 right 278.36. The selection of times is the same as indicated by the arrows
 in Fig. \ref{f39}. }
\label{f44}
\end{figure*}

 First, up to about 10 \tref
 we have a fast binary mass segregation phase, in which the total number of
 binaries does not change much, since the time scale for their destruction
 and ejection is still too long, but as the binary distribution flattens out
 into core and halo, the maximum of the distribution becomes shallower.
 The effect can be clearly seen in the difference between
 snapshot 0 and 1, and by the pronounced maximum of central escape speed
 reached at time point 1 in Fig.~\ref{f39}. Subsequently there follows a 
 phase of rapid binary destruction, as can be seen in Fig.~\ref{f37} up
 to about 30 \tref; remarkably from the snapshot 2 it can be seen that
 preferrably the low energy binaries (up to a few tens of $kT$)
 are being destroyed, while the high energy binaries are not affected
 much. Here we see the 4b encounters being effective: with a
 high probability the low energy binaries are undergoing a close encounter
 with a moderate or high energy binary, leading to the destruction of the
 first and hardening of the second binary. Depending on the energy of the
 second binary, escape of single stars only or of single stars
 and the hard binary may happen;
 if in the beginning there are still enough binaries of moderate
 binding energy the dominant effect is just heating of the surrounding
 core by 3b and 4b encounters. Therefore we observe
 a quasi-stationary binary burning phase lasting up to about 60 \tref,
 during which the ratio of binary and single star density is kept nearly
 constant in the core (Fig.~\ref{f38}), with a gradually expanding
 core radius of the single stars (Fig.~\ref{f36}). At about 60
 \tref the reservoir of binaries to destroy and heat the core
 is substantially depleted, the binaries have been reduced in number by 60\%.

\begin{figure*}
\vskip 19truecm
\includegraphics{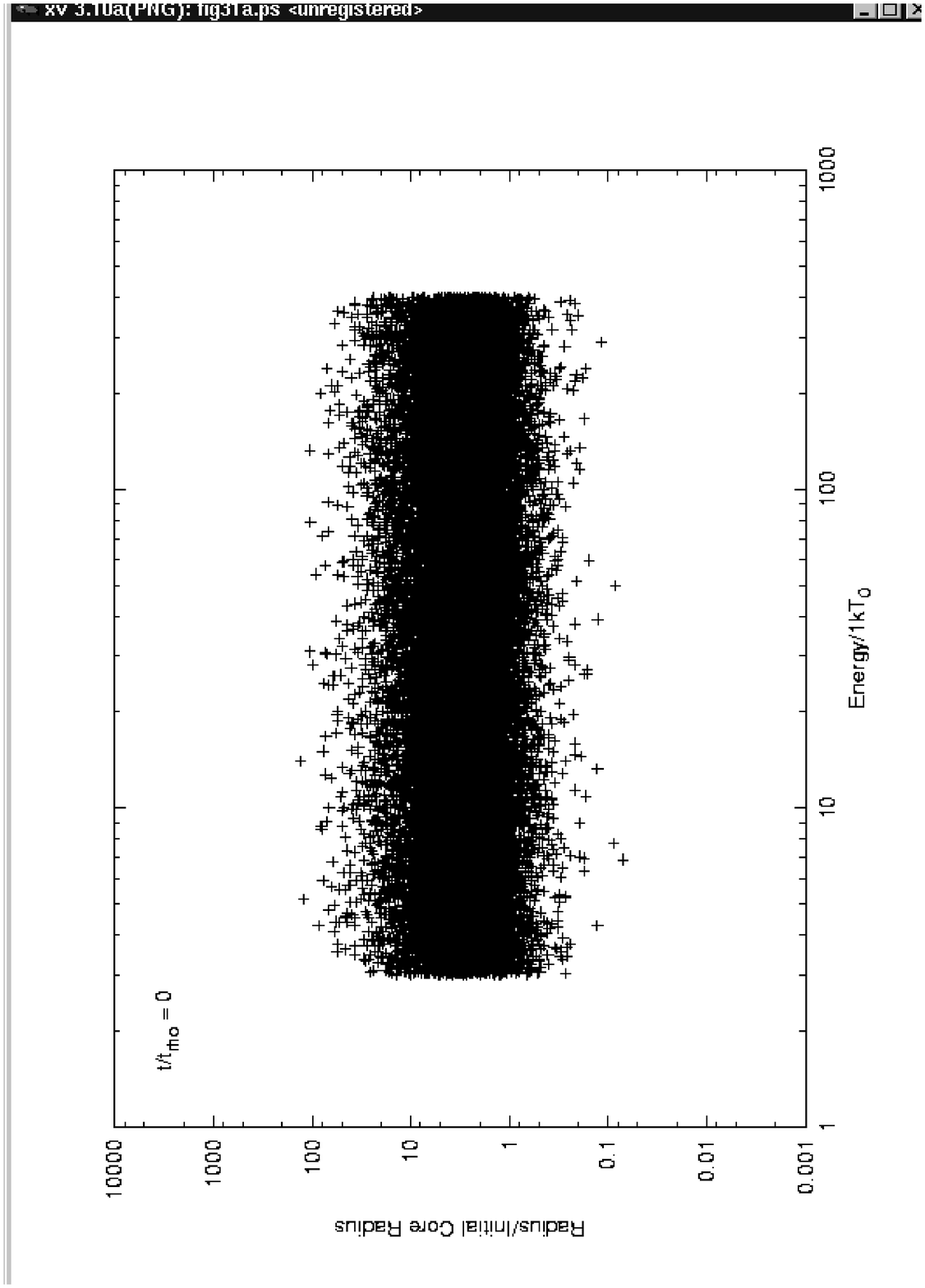}
\includegraphics{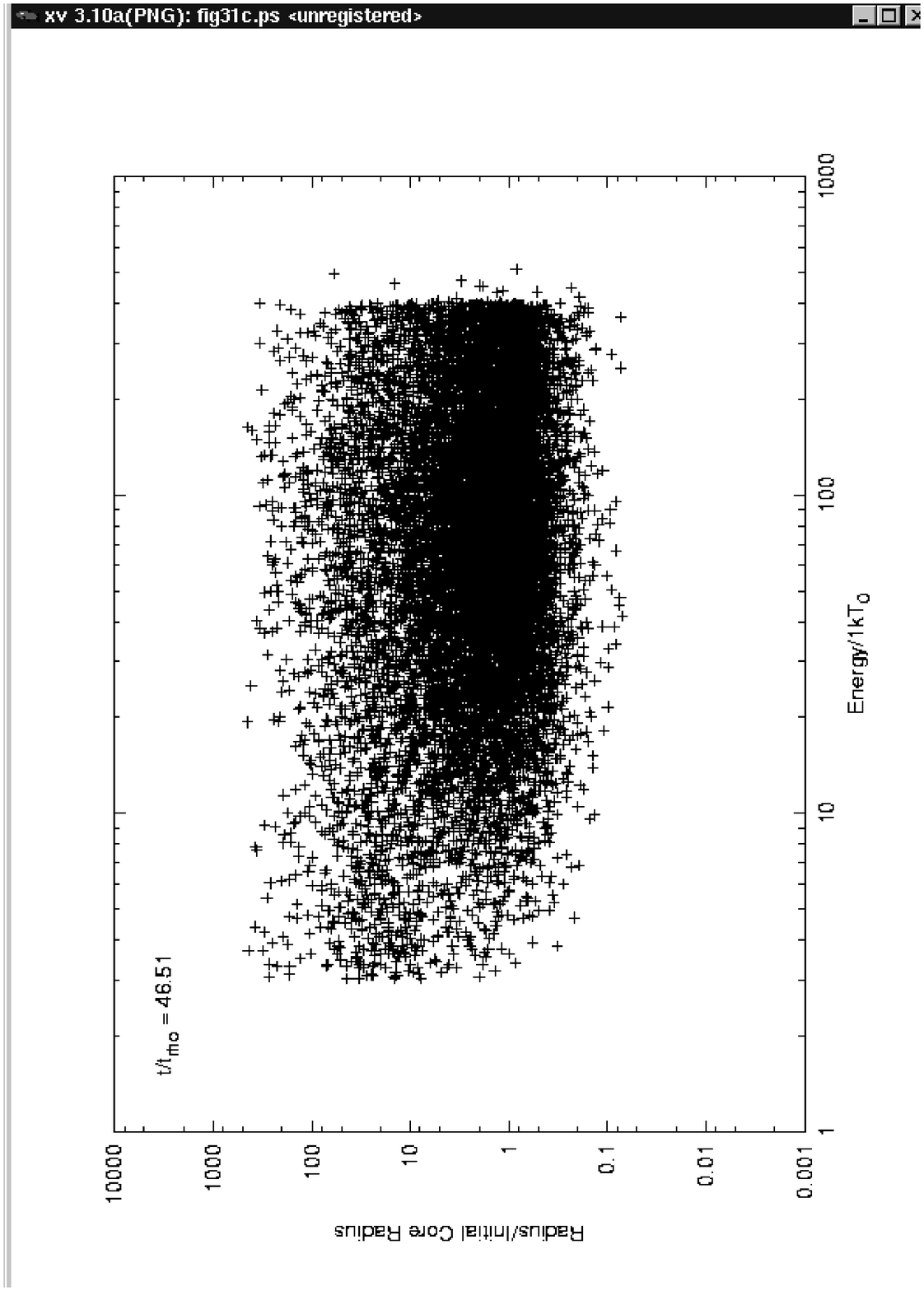}
\includegraphics{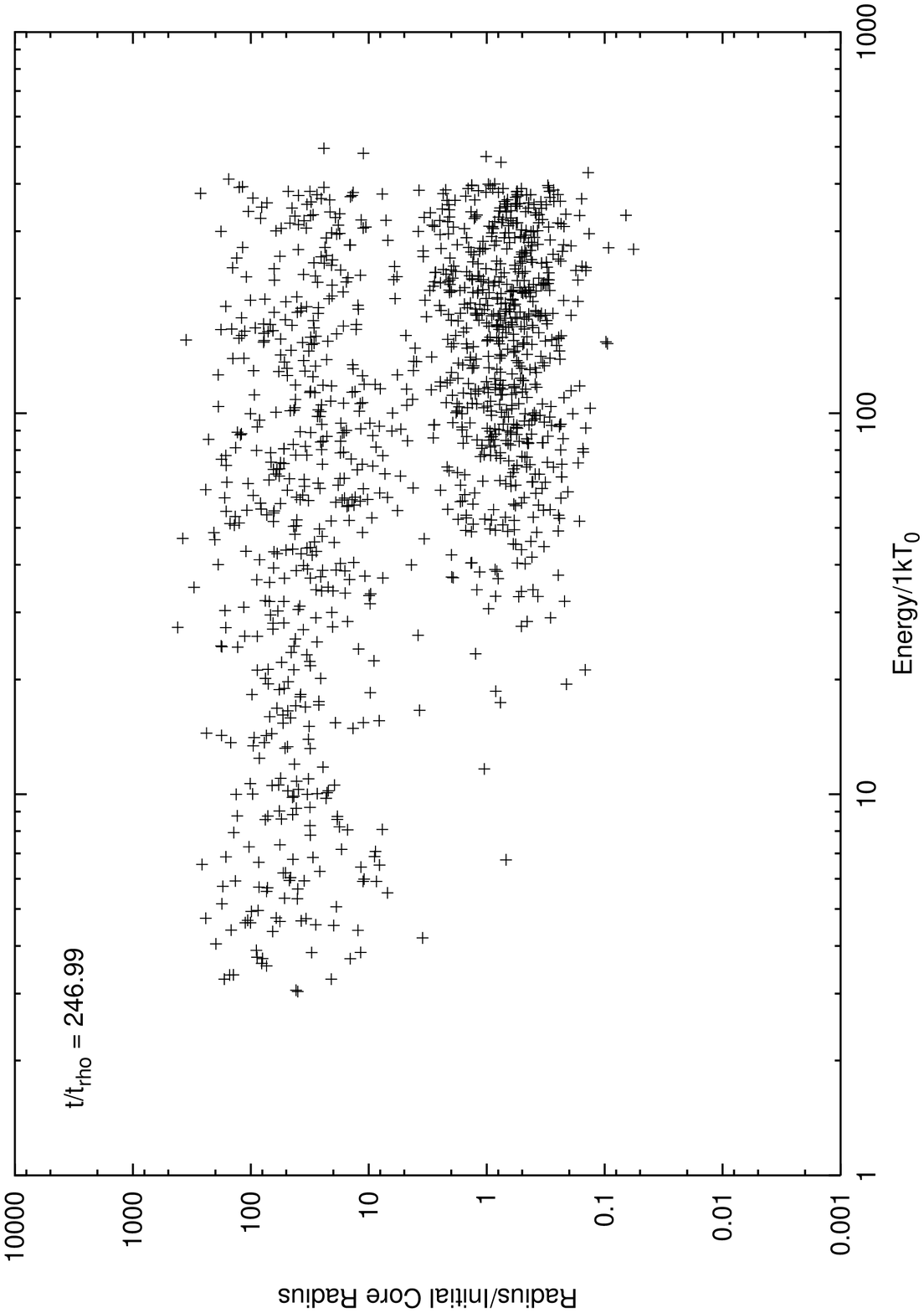}
\includegraphics{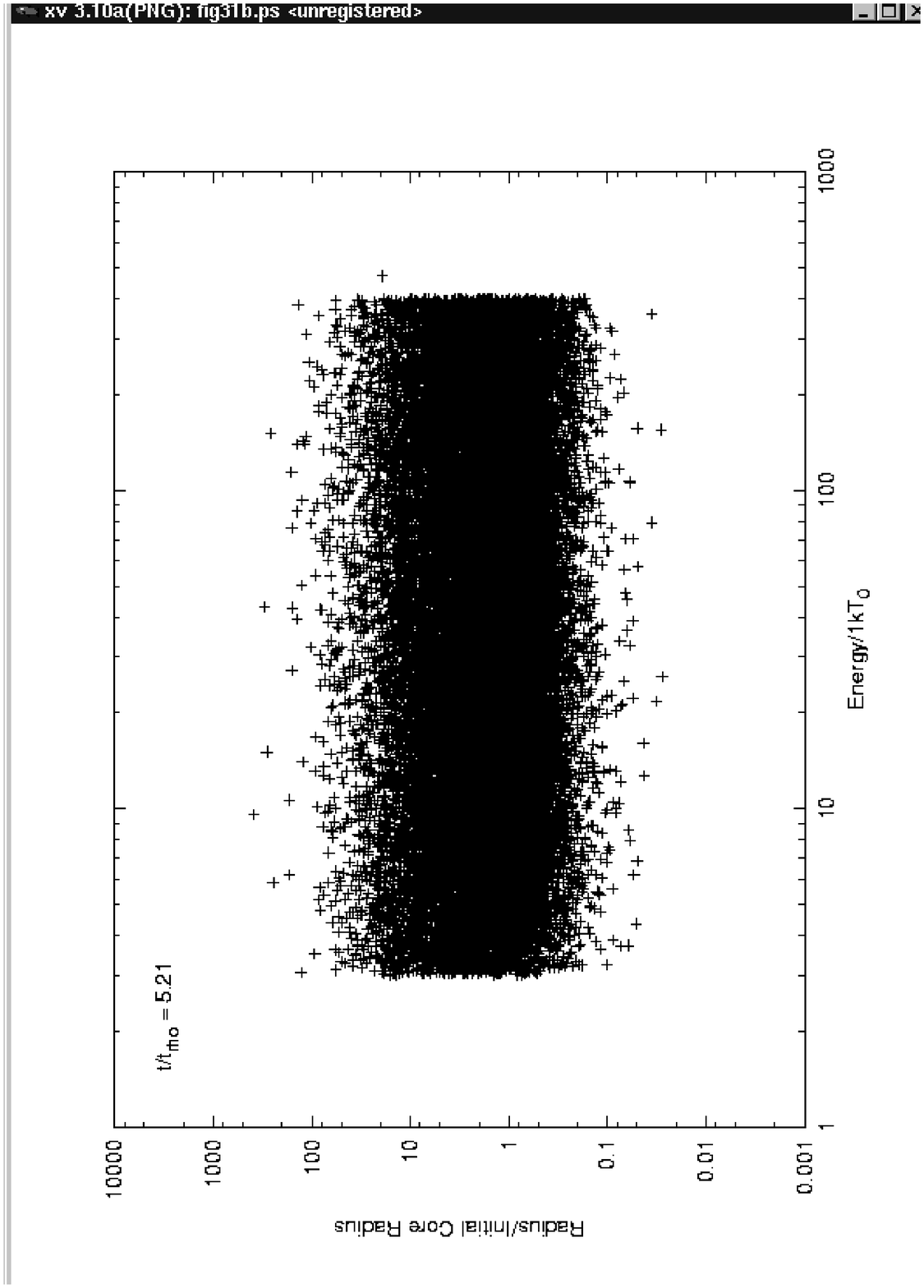}
\includegraphics{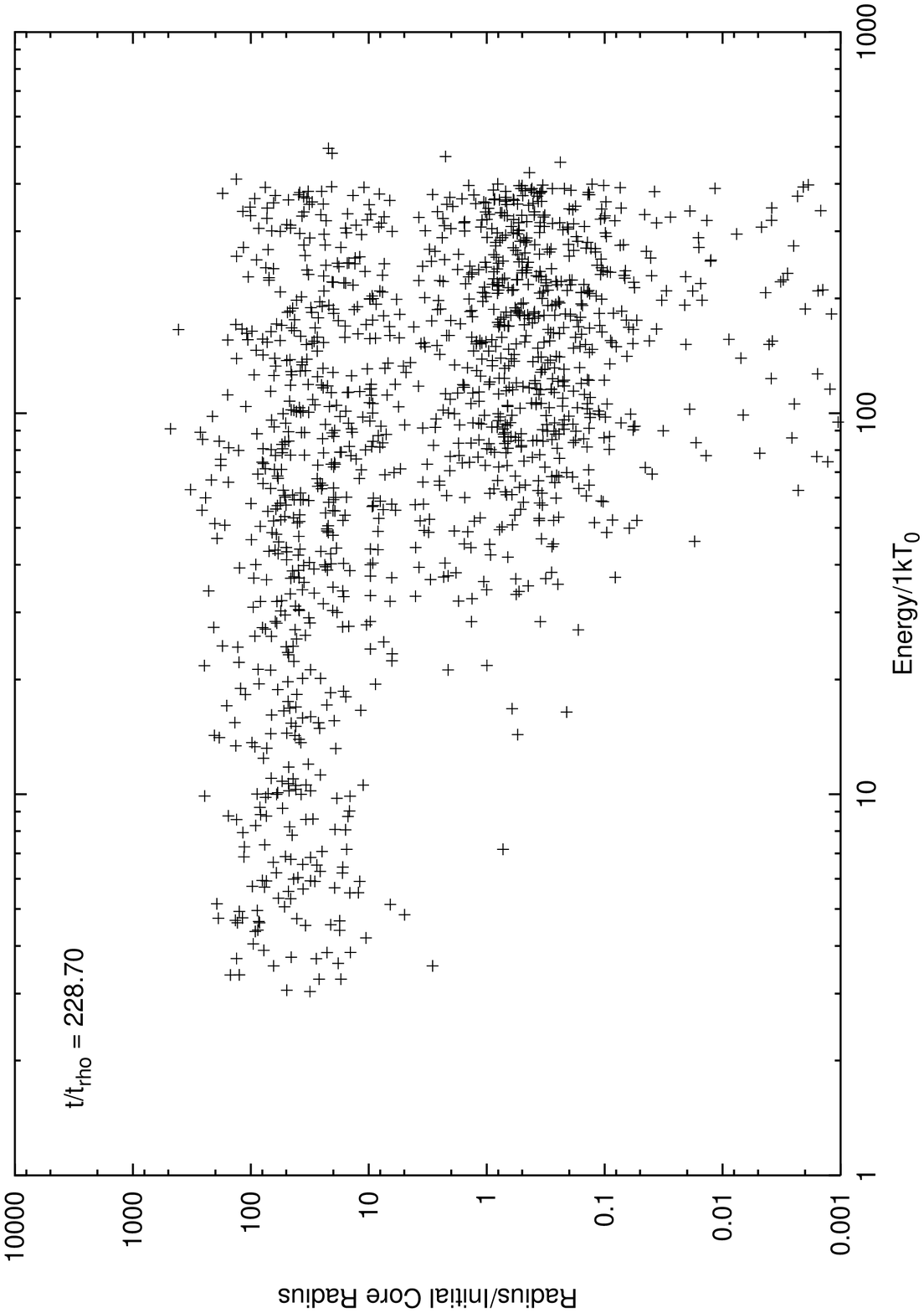}
\includegraphics{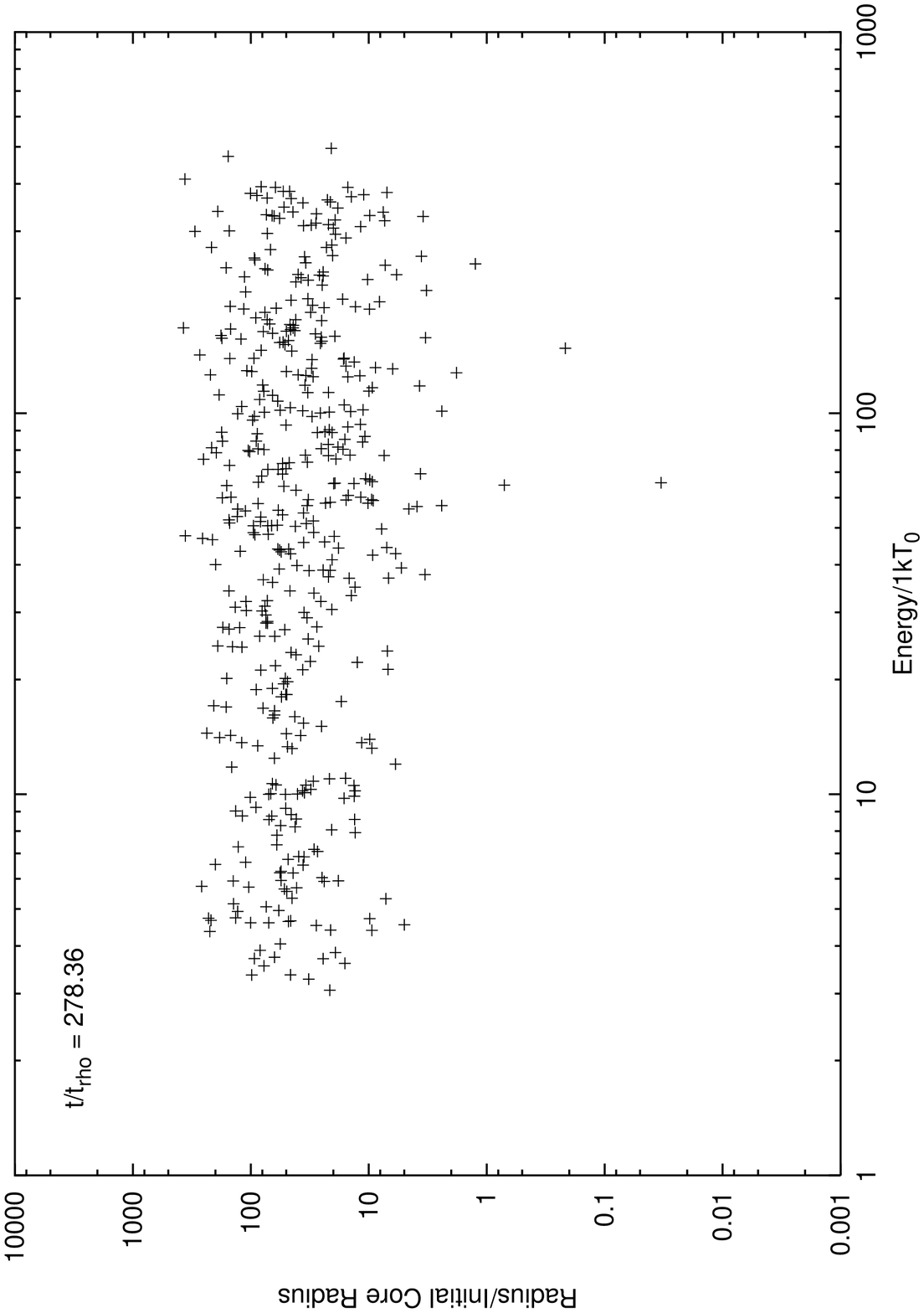}

\caption{The same data as in Fig. \ref{f44}, but in a 2D projection of
 the individual data of each binary (being represented by a cross) onto
 the energy-radius plane, all units as in Fig. \ref{f44}.}
 \label{f45}
\end{figure*}
 The core starts its first attempt to collapse gravothermally,
 which is visible as a transient recollapse of the core radius,
 a shoulder in the binary mass fraction ($dM/dt$ of the binaries
 becomes significantly smaller temporarily), and an increase of
 the ratio of binary to single star density, as a consequence of
 mass segregation, all three effects clearly visible between
 50 and 70 \tref in Figs.~\ref{f36} to \ref{f40}. However,
 there is still a large enough fraction of the initially 
 30.000 binaries present to
 halt core collapse at a higher level of binary to single
 star density, and cause a second quasi-stationary phase until
 about 150 \tref. The process repeats itself, but now
 at a small enough binary number that the single stars can
 start to undergo the first pronounced oscillatory peaks
 (small core radius). Gravothermal oscillations follow, which are
 pronounced at the time points 3 (maximum density), 4 (expanded
 stage) and 5 (terminal point of our model). It is interesting
 to note, that the gravothermal oscillations are not only visible
 in the core radius and Lagrangian radii of the single stars. They are
 also visible in
 the binary density, and in the {\em ratio} of binary to
 single star density (see Figs. \ref{f39} and \ref{f40}). These features,
 including amplitude variations over many orders of magnitude, multi-peak
 structure with period doublings at the maxima, and long inactive
 expanded phases are clear and the same as observed in GGCM91 and
 other standard models of gravothermal oscillations. From
 Fig.~\ref{f44}, time points 3 and 4 we can see, that in the
 collapsed phase (time point 3) some binaries are very deep in the core,
 while in the expanded phase (time point 4) the core is void of
 binaries, no activity taking place. Most interestingly at the
 final time point 5, we have reached a situation where all
 binaries are very far outside in parking orbits. We note, however,
 that from the studies of MC runs (MC5 and MC5Q, subsection 3.2) the reason,
 why there are so many binaries in parking orbits is not yet
 clear. In contrast to MC runs, however, here most binaries
 in parking orbits are still primordial and have not been created
 in 3b encounters in high density phases. Therefore we
 believe that the existence of many binaries in such orbits is
 real, though our model may overestimate it.
 In any case, such binaries decouple from the evolution
 of the rest of the system, which will behave more like a
 single star cluster with 
 gravothermal oscillations. Binaries left
 over in the outer halo may be regarded as fossil records of the early
 primordial binary generation. They will be slowly depleted
 by 3b and 4b interactions when they enter the core due to
 relaxation processes. Note that in phases
 3 to 5 the scale of the binary distribution in Fig.~\ref{f44}
 has been amplified by a large factor; the absolute frequency
 (and total number) of binaries are one to two orders of magnitude
 smaller than at time points 0 to 2, as can be deduced directly
 from Fig.~\ref{f37}

\begin{figure}
 \psfig{figure=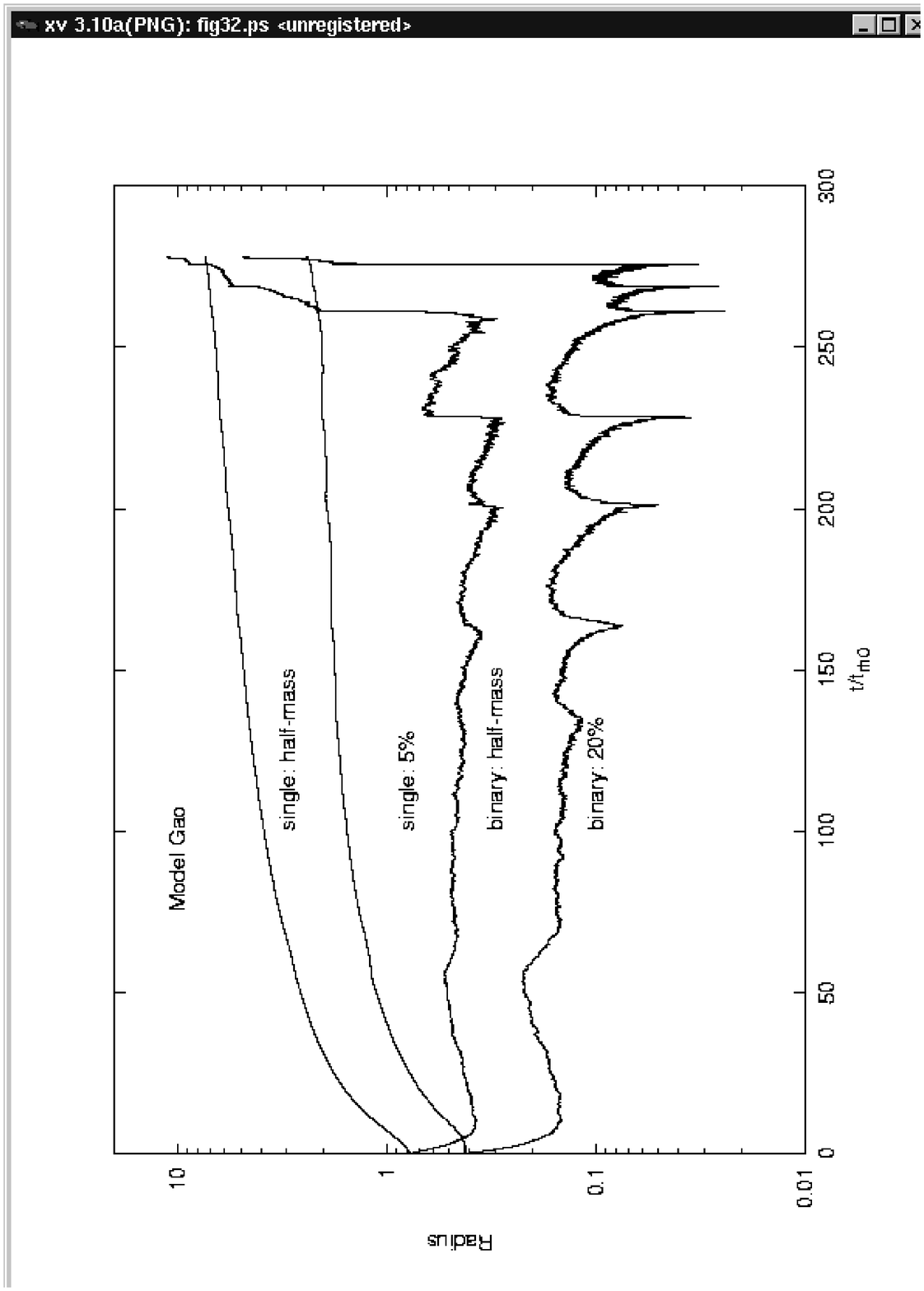,height=5.5cm,width=8.5cm,angle=-90}
 \caption{Evolution of Lagrangian radii containing $50\%$ and $5\%$ of the
 mass of single stars and $50\%$ and $20\%$ of the mass of
 binaries as a  function of time.}
 \label{f41}
 \end{figure}

\begin{figure}
 \psfig{figure=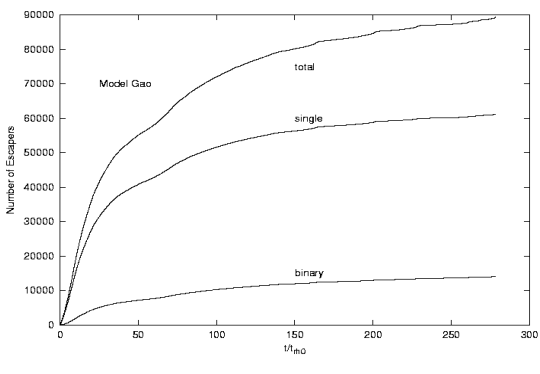,height=5.5cm,width=8.5cm,angle=-90}
 \caption{Number of escaping single stars, binaries and the total number
 of escapers as a function of time.}
 \label{f42}
 \end{figure}

\begin{figure}
 \psfig{figure=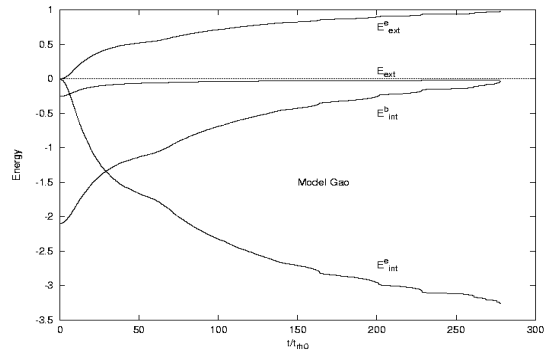,height=5.5cm,width=8.5cm,angle=-90}
 \caption{Energy balance as a function of time. The four different
 contributions are described in the text, and are the same as used
 for Figs. \ref{f5} and \ref{f6} for Heggie's runs.}
 \label{f43}
 \end{figure}

 Figs.~\ref{f41}, \ref{f42}, and \ref{f43} show the evolution of
 selected Lagrangian radii for binaries and singles, the number
 of escaping stars in binaries, singles, and total (counting
 binaries as two stars), and the total energy budget, in the
 same way as discussed for Heggie's models. Due to their
 stronger central concentration the binaries take part
 in the gravothermal oscillations with more than 50\% of
 their total mass; at the end the final complete
 removal of binaries from the core can be seen in Fig.~\ref{f41}. 
 Figs.~\ref{f42} and
 \ref{f43} show that the initial binary destruction phase is
 accompanied also by a heavy loss of binaries with high binding
 energy, as can be seen from the strong drop of $E^e_{\rm int}$.
 In the following phase the rate of escape becomes slower,
 but the mechanism is the same, energy bound in binaries is
 carried away by escapers. Note the signature of the oscillations
 in the binary escaper energy at late times, and the enormous
 amount of energy exchanged via the binaries, as compared to
 the initial and final total external energy of the star cluster,
 which starts at the standard value of -0.25. Remarkably
 we lost in total at the end of our simulation about 90.000 stars
 escaped altogether, which is one quarter of the initial number of stars.
 This number would be even larger if escapers due to relaxation
 effects of single stars were allowed.
 Since we lose only some 15.000 binaries by escape (Fig.~\ref{f42}),
 it can be concluded that another 15.000 binaries have been
 destroyed by close 4b encounters, and, more interestingly,
 at least 30.000 single stars have escaped which do {\em not}
 originate from one of the primordial binaries, assuming that
 the maximum of all destroyed binaries led to two single escapers,
 which is an upper limit only, and neglecting possible exchange
 reactions. In other words, per binary destroyed or ejected about three
 single stars escaped on average. 
 Looking at Fig.~2 of GGCM91 they find an increased number of single stars
 during the first evolutionary phase, which cannot be seen in
 our results. We think that this difference is an artifact of the
 GGCM91 runs, due to their rather artificial procedure to select
 binaries from energy bins for close encounters, while in our
 paper the proper encounter probabilities are used. Thereby we
 destroy less quickly the soft binaries and find a higher fraction
 of escaping single stars from the beginning, so in our results
 the number of bound single stars decreases. Another reason why
 GGCM91 find more bound single stars is that they can not take into
 account the external energy of the destroyed binary in a 4b
 encounter, as a further contribution to the possibly escaping 
 single stars in excess of the recoil energy obtained by hardening
 the harder of the two binaries (though the latter will in
 most cases be the dominant contribution).

 Hence, we can explain the time scale differences between our
 and GGCM91 models; we keep a larger number of soft or intermediate
 binaries in the system which are able to supply sufficient binary
 heating, by single stars originating from
 4b encounters and remaining in the core. These
 heating reactions support our long initial binary burning phase,
 which takes place in two phases (as discussed above) until
 150 \tref. Opposed to this, the soft and intermediate binaries
 in GGCM91 are quickly destroyed, leading to a much stronger collapse
 of the single star and the remaining relatively hard
 binaries, and further acceleration
 of the recoil and escape process. Consistently one could
 see that the gravothermal oscillations in GGCM91 already
 begin at a fraction of binaries left of about 30 to 40\%,
 while in our case we have a binary depletion to 10\% until
 the oscillations start. Since they destroy the weak binaries too quickly
 this difference in evolution between the models occurs, although
 the time evolution of the total binary number is not so different.
 All this leads to a time,
 to begin gravothermal oscillations, which is about a factor of 3
 shorter than in our models. This is a big difference, which
 however, can be fully explained by a much faster binary destruction
 in their model. We attribute this to the rather artificial procedure
 by which GGCM91 decide whether a close 4b encounter takes place.
 
 \section{Conclusions and discussion}

 The new approach outlined in Paper I (Spurzem \& Giersz 1996) to follow, 
 in the 
 Monte Carlo manner, the individual formation and evolution of binaries in an 
 evolving point-mass cluster was successfully extended to the fully 
 self-consistent treatment of relaxation and close three- and four-body
 (abbreviated 3b and 4b)
 encounters for a substantial number (typically a few percent of the initial 
 number of stars) of binaries and a realistic total number
 of stars. We use a standard anisotropic gaseous model (Louis \& Spurzem 1991,
 Spurzem 1994), 
 describing the single stars, and the Monte Carlo technique (Giersz 1998) 
 to model, in a stochastic way, the binary subsystem. The aim of this 
 paper was to test the reliability of the new hybrid code by comparing its 
 results with the data available in the literature for single mass systems
 (full Monte Carlo model - Giersz 1998), and for systems with 
 substantial primordial binary population ($N$-body models - HA92 and 
 Fokker-Planck models - GGCM91). Also we want  to show, that our model is able
 to handle large $N$ systems with a large binary number by a reasonable
 effort, but keeps full self-consistency and detailed information. It will
 be possible in the near future, after some more real physics is
 included (such as stellar evolution, finite stellar radii, tidal fields,
 and a mass spectrum), to provide unique model data for comparisons with
 the expected wealth of observational data for globular clusters.
 
 The features of our MC5 and MC5Q runs for isolated, single mass systems
 consisting of $10^5$ stars with only strong 3b (MC5) and both 3b and 4b (MC5Q)
 interactions, respectively, are in a good agreement with the full Monte Carlo 
 model (Giersz 1998). The gravothermal oscillations are the most pronounced
 features of these runs. They show very large amplitude in central density, 
 long expanded (inactive) phases, and several discrete
 oscillation frequencies which are observed in standard
 gaseous or Fokker-Planck models of gravothermal oscillations (Bettwieser
 \& Sugimoto 1984, Heggie \& Ramamani 1989, Cohn, Hut \& Wise 1989,
 Breeden et al. 1994).
 Also rapid changes in the phase of the oscillations occur, which agree
 with other stochastic Fokker-Planck, $N$-body and Monte Carlo models 
 (Takahashi 
 \& Inagaki 1991, Makino 1996 and Giersz 1998). As can be expected, the 
 expansion phases of the system are powered by the themperature inversion in 
 the core, not by the energy generated by binaries in interaction with singles 
 and other binaries. In the large expansion phases there is practically no 
 binary activity, as expected. The binary distribution in energy-radius space 
 of the cluster is clearly bimodal. Binaries with high binding energies and in 
 orbits not entering the core and extended far into the halo form one 
 group. Binaries with a wide range of binding energies and orbits entering 
 the core form the second group. From models with artificially
 suppressed 4b encounters we find that they cause
 a wider distribution of binary binding energies (mainly towards
 larger binding energies) and a somewhat less pronounced bimodal binary
 distribution in a diagram showing the binary distribution
 in radius and binding energy. Our models exhibit a
 rather large number of binaries in parking orbits (orbits which do not
 reach the core, and have apocentre very far out in the halo), as compared
 to full Monte Carlo (Giersz 1998) and direct $N$-body models (Spurzem
 \& Aarseth 1996). The reason is unclear, but see a discussion in
 subsections 3.2 and 3.3. In any system with even a slight external
 tidal field such binaries will be removed very fast.
 
 For runs with primordial binaries we discuss three different cases.
 First, we use an artifical model with two-components, one consisting of
 single, the other of binary stars, which only interact by relaxation
 with each other and with the single (all close 3b and 4b
 encounters were suppressed). We check the rate of mass segregation
 and core collapse time by comparison with a continous two-component
 anisotropic gaseous model (Louis \& Spurzem 1991, Spurzem 1994),
 just in the same way as HA92 checked their $N$-body models by comparison
 with one of Heggie's gaseous models. We find that our stochastic Monte
 Carlo model of the binary component provides a good match with
 the expectation from the gaseous model. While in Paper I we only checked
 the dynamical friction of one binary in a system of single stars, this
 is a more significant test of the self-consistent treatment of relaxation 
 in our model. 

 Second, a set of four models corresponding to the largest
 published full $N$-body models with primordial binaries by Heggie 
 \& Aarseth (1992, HA92) was performed. We find a good agreement
 of our stochastic Monte Carlo model for most features, as they were
 published in HA92. The remaining differences regarding the
 binary number evolution, central potential, and fraction of hard
 binaries in the system between their and our models
 can be related to stochastic variations, as
 they usually occur in individual
 Monte Carlo realizations of the system.
 
 Third, and as a final goal of this paper,
 we have performed a self-consistent model of 30.000 binaries evolving 
 internally and externally,
 surrounded by a live star cluster of 300.000 single stars,
 undergoing mass segregation, core collapse, and finally large amplitude
 gravothermal oscillations. The binaries during all these evolutionary phases
 are subject to binary destructions and ejections
 by close 3b and 4b encounters. 
 This model is called Gao's run, to remind the
 reader of the pioneering work of Gao et al. (1991, GGCM91), which performed
 for the first time a self-consistent two-component Fokker-Planck model of
 binaries and single stars. Our aim was, in this paper, to redo this model
 as much as possible and reasonable, keeping most of the assumptions, such
 as the 3b and 4b interaction cross sections, the assumption of
 single mass (all binaries consist of mass components equal to the
 single stars), and the assumption of an isolated point-mass
 system. However, in addition to their model, we exploit in this framework the
 strengths of our stochastic Monte Carlo model, such that we are able to
 follow the individual evolution of internal and external parameters of any
 binary in nearly the same detail as in an $N$-body model, provide snapshots
 of the position and binding energy of all the binaries at crucial times
 of the evolution (first mass segregation, binary heating and destruction
 phase, gravothermal oscillation phase), which can be seen by the interested
 reader as a movie under
 {\parindent=0pt
 
 {\tt ftp://ftp.ari.uni-heidelberg.de/ 

  \hfill pub/spurzem/movie?.mpg}

 where ``?'' refers to 1 or 2, depending whether one wants to see} 
 the movie in the
 style of Fig.~\ref{f44} or \ref{f45}. Also we do not use the rather artificial
 procedure of picking binaries in the energy bins for close encounters as
 used by GGCM91, because in their model no better procedure could
 be found. Since we know positions of binaries at any time, we can select
 proper probabilities for close encounters of the 3b and 4b
 kind to occur and find a much slower destruction rate of soft and intermediate
 binaries than GGCM91. As a consequence the quasistationary binary burning
 phase extends over a time in our model (up to 150 \tref) which is about three
 times larger than the corresponding time in GGCM91. After that,
 gravothermal oscillations start, and at about
 280 \tref all primordial binaries left the core and its vicinity.
 Only a fossil collection of some 400 binaries can be found in the
 very outer halo, as a final trace of the initial primordial binary
 population. During the maximum density phase new binaries
 start to be formed by 3b interactions.
 Of the initially 30.000 binaries, 15.000 were ejected by
 3b and 4b encounters, and another 15.000 destroyed by
 4b encounters. The single stars originating from such binary
 destructions are partly heating the cluster, partly escaping, depending
 on the binding energy of the harder binary in such encounters. On
 average we find that for any binary, whether destroyed or ejected,
 about three single stars are ejected.
 Qualitatively we find the same phases as GGCM91, but our evolution
 time scales are a factor of 3 longer. Though the evolutionary
 time scale is much longer than the Hubble time, we think our models
 are still (already on the present idealized level) able to provide some
 insight into the physical evolution of large star clusters with many
 primordial binaries, at least if accepting the point mass approximation. 
 Note, that the inclusion of more realism (see
 below) according
 to previous experiences always shortens the evolutionary lifetimes
 of clusters, and stellar mass loss will change the cluster's and the
 individual binaries evolution dramatically. Nevertheless we consider
 our work as an important and indispensable step towards such
 models with very large binary and single star numbers.

 Admittedly our model is still not realistic for a globular cluster in
 many respects. The evolution of hard binaries during close encounters
 is known to be {\em very} different from that of equal mass stars;
 stellar evolution and finite size effects will dramatically alter
 the channels through which primordial binaries are processed
 (see Hut, McMillan \& Romani 1992). Tidal fields may strip the 
 outer areas of the cluster, especially those where we find our
 remaining binaries at the end. Nevertheless we stress, that in
 our model we have achieved a breakthrough in the homogeneous, self-consistent
 treatment of very many binaries in a large star cluster. We model
 the cluster evolution during
 0.8 million $N$-body time units, which are some 260.000 initial half-mass
 crossing times, without either using special hardware
 or other than the fundamental assumptions
 of spherical symmetry and dominance of small angle two-body encounters
 (which are very robust assumptions in large $N$ systems).
 Such a job is enormous, even for the next generation of
 Petaflop computers if done by a direct $N$-body model. In this sense
 our model has proven its usefulness, uniqueness, and pioneering
 capabilities, which gives credit to those people, who have been
 developing the Monte Carlo method
 (H\'enon 1971, Spitzer 1975, \Stodolkiewicz 1982, 1986), and
 it shows the relevance of the new Monte Carlo models developed
 recently (Giersz 1996, 1998).
 
 The yet missing effects (multi-mass, stellar evolution, tidal fields,
 more detailed cross-sections, obtained by direct modelling) will
 be included into our models for the near future, and do not pose a
 fundamental challenge. But in order to do the proper astrophysical
 study, we rather prefer to do it step by step, connecting our models
 with Heggie's, Giersz's and Gao's runs, discuss their reliability,
 agreement and disagreement and the physical reasons of it, to understand
 what goes on in our model, as compared to just start with one big model
 containing everything.

 \section*{Acknowledgements}
This work was supported in part by the Polish National Committee for Science 
Research under grant 2 P03D 022 12. Financial Support by German Science
 Foundation under grant no 436 POL 17/18/98, and
 436 POL 18/3/99 is gratefully acknowledged. R.Sp. and CAMK wish to thank
 CAMK and ARI for their hospitality during mutual research visits.
 We thank Sverre Aarseth for
 a very careful reading and criticism of the manuscript.


\begin{thebibliography}{}
 \bibitem[]{} Aarseth S.J., 1985, in Brackbill J.U.,
   Cohen B.I., eds, Multiple time scales, Academic Press, Orlando,
      p. 378
 \bibitem[]{} Aarseth S.J., 1996, in Hut P., Makino J., eds,
 Proc. IAU Symp. 174, Dynamics of Star Clusters, Reidel, Dordrecht, p. 161
 \bibitem[]{} Aarseth S.J., 1999, CeMDA, 73, 127
 \bibitem[]{} Aarseth S.J., Heggie D.C., 1998, MNRAS, 297, 794
 \bibitem[]{} Bettwieser E., Sugimoto D., 1984 MNRAS, 208, 493
 \bibitem[]{} Breeden J.L., Cohn H.N., Hut P., 1994, ApJ, 421, 195
 \bibitem[]{} Cohn H., 1980, ApJ, 242, 765
 \bibitem[]{} Cohn H., Hut P., Wise M.W., 1989, ApJ, 342, 814
 \bibitem[]{} Casertano S., Hut P., 1985, ApJ, 298, 80
 \bibitem[]{} Einsel Ch., Spurzem R., 1999, MNRAS, 302, 81
 \bibitem[]{} Gao B., Goodman J., Cohn H., 
    Murphy B., 1991, ApJ, 370, 567, GGCM91
 \bibitem[]{} Giersz M., 1996, in Hut P., Makino J., eds,
 Proc. IAU Symp. 174, Dynamics of Star Clusters, Reidel, Dordrecht, p. 101
 \bibitem[]{} Giersz M., 1998, MNRAS, 298, 1239
 \bibitem[]{} Giersz M., Heggie D.C., 1994a, MNRAS, 268, 257, GHI
 \bibitem[]{} Giersz M., Heggie D.C., 1994b, MNRAS, 270, 298, GHII
 \bibitem[]{} Giersz M., Heggie D.C., 1997, MNRAS, 286, 709
 \bibitem[]{} Giersz M., Spurzem R., 1994, MNRAS, 269, 241, GS
 \bibitem[]{} Heggie D.C., 1984, MNRAS, 206, 179
 \bibitem[]{} Heggie D.C., Aarseth S.J., 1992, MNRAS, 257, 513, HA92
 \bibitem[]{} Heggie D.C., Giersz M., Spurzem R., Takahashi K., 1998, in
 Andersen J. (ed), Highlights of Astronomy Vol. 11. Kluwer, Dordrecht.
 (Preprint astro-ph/9711197)
 \bibitem[]{} Heggie D.C., Mathieu R.M. 1986, in Hut P., McMillan S.L.W.,
  eds, The Use of Supercomputers in Stellar Dynamics. Springer-Verlag,
  Berlin, p. 233
 \bibitem[]{} Heggie D.C., Ramamani N., 1989, MNRAS, 237, 757
 \bibitem[]{} H\'enon M., 1971, Ap\&SS, 14, 151
 \bibitem[]{} H\'enon M., 1975, in A. Hayli, ed, Proc. IAU Symp. 69,
   Dynamics of Stellar Systems,
   Reidel, Dordrecht, p. 133
 \bibitem[]{} Hut P., et al., 1992, PASP, 104, 981
 \bibitem[]{} Hut P., McMillan S., Romani R.W., 1992, ApJ, 389, 527
 \bibitem[]{} Inagaki S., Hut P., 1988, in Valtonen M.J., ed, Proc. IAU
 Coll. 96, The Few Body Problem, Kluwer, Dordrecht, p. 319
 \bibitem[]{} Kroupa P., 1995, MNRAS, 277, 1491
 \bibitem[]{} Lee H.M., Fahlman G.G., Richer H.B., 1991, ApJ, 366, 455
 \bibitem[]{} Louis P.D, Spurzem R., 1991, MNRAS, 251, 408
 \bibitem[]{} Makino J., 1996, in Hut P., Makino J., eds, 
 Proc. IAU Symp. 174, Dynamics of Star Clusters, Reidel, Dordrecht, p. 141
 \bibitem[]{} Makino J., Aarseth S.J., 1992, PASJ, 44, 141
 \bibitem[]{} Makino J., Taiji M., 1998, Scientific simulations with
 special purpose computers, Chichester, Wiley
 \bibitem[]{} Makino J., Taiji M., Ebisuzaki T., Sugimoto D., 
 1997, ApJ, 480, 432
 \bibitem[]{} McMillan S.L.W., Engle E.A., 1996, in Hut P., Makino J., eds,
 Proc. IAU Symp. 174, Dynamics of Star Clusters, Reidel, Dordrecht, p. 379
 \bibitem[]{} McMillan S.L.W., Hut P., Makino J., 1990, ApJ, 362, 522
 \bibitem[]{} McMillan S.L.W., Hut P., Makino J., 1991, ApJ, 372, 111
 \bibitem[]{} McMillan S.L.W., Hut P., 1994, ApJ, 427, 793
 \bibitem[]{} McMillan S.L.W., Hut P., 1996, ApJ, 467, 348
 \bibitem[]{} Mikkola S., 1983a, MNRAS, 203, 1107
 \bibitem[]{} Mikkola S., 1983b, MNRAS, 205, 733
 \bibitem[]{} Mikkola S., 1984a, MNRAS, 207, 115
 \bibitem[]{} Mikkola S., 1984b, MNRAS, 208, 75
 \bibitem[]{} Portegies Zwart S.F., Hut P., Makino J., McMillan S.L.W.,
 1998, A\&A, 337, 363
 \bibitem[]{} Sigurdsson S., Phinney E.S., 1995, ApJS, 99, 609
 \bibitem[]{} Spitzer L., 1975, in Hayli A., ed, 
 Dynamics of Stellar Systems, Reidel: Dordrecht, p.3
 \bibitem[]{} Spitzer L., 1987, Dynamical Evolution of Globular
  Clusters, Princeton Univ. Press, Princeton
 \bibitem[]{} Spurzem R., 1992, in Klare G., ed, 
 Reviews of Modern Astronomy 5, Springer Verlag Berlin Heidelberg, p.161
 \bibitem[]{} Spurzem R., 1994, in Pfenniger D., Gurzadyan V.G., eds,
 Ergodic Concepts in Stellar Dynamics, Springer-Vlg.,
 Berlin, Heidelberg, p. 170
 \bibitem[]{} Spurzem R., 1996, in Hut P., Makino J., eds, Dynamics
 of Star Clusters, Proc. IAU Symp. No. 174, p.111
 \bibitem[]{} Spurzem R., 1999, in Riffert H., Werner K. (eds),
 Computational Astrophysics,
 The Journal of Computational and Applied Mathematics (JCAM) 109,
 Elsevier Press, Amsterdam, p. 407
 \bibitem[]{} Spurzem R., Aarseth S.J., 1996, MNRAS, 282, 19
 \bibitem[]{} Spurzem R., Giersz M., 1996, MNRAS, 283, 805, Paper I
 \bibitem[]{} Spurzem R., Takahashi K., 1995, MNRAS, 272, 772
 \bibitem[]{} \Stodolkiewicz J.S., 1982, Acta Astronomica, 32, 63
 \bibitem[]{} \Stodolkiewicz J.S., 1985, in Goodman J., Hut P., eds, Proc.
   IAU Symp. 113, Dynamics of Star Clusters, Reidel, Dordrecht, p. 361
 \bibitem[]{} \Stodolkiewicz J.S., 1986, Acta Astronomica, 36, 19
 \bibitem[]{} Sugimoto D., Chikada Y., Makino J., Ito T.,
    Ebisuzaki T., Umemura M., 1990, Nature, 345, 33
 \bibitem[]{} Takahashi K., 1995, PASJ, 47, 561
 \bibitem[]{} Takahashi K., 1996, PASJ, 48, 691
 \bibitem[]{} Takahashi K., 1997, PASJ, 49, 547
 \bibitem[]{} Takahashi K., Inagaki S., 1991, PASJ, 43, 589
 \bibitem[]{} Takahashi K., Lee H.M., Inagaki S., 1997, MNRAS, 292, 331

 \end{thebibliography}
 \end{document}